%Paper: hep-th/9310026
%From: gregory moore <moore@castalia.physics.yale.edu>
%Date: Wed, 6 Oct 93 22:45:38 -0400
%Date (revised): Sun, 10 Oct 93 17:54:39 -0400

\input harvmac.tex

\def\IL{\relax{\rm I\kern-.18em L}}
\def\IP{\relax{\rm P\kern-.18em P}}
% Poor man's Blackboard Bold characters often used :
\def\inbar{\,\vrule height1.5ex width.4pt depth0pt}
\def\IB{\relax{\rm I\kern-.18em B}}
\def\IC{\relax\hbox{$\inbar\kern-.3em{\rm C}$}}
\def\ID{\relax{\rm I\kern-.18em D}}
\def\IE{\relax{\rm I\kern-.18em E}}
\def\IF{\relax{\rm I\kern-.18em F}}
\def\IG{\relax\hbox{$\inbar\kern-.3em{\rm G}$}}
\def\IH{\relax{\rm I\kern-.18em H}}
\def\II{\relax{\rm I\kern-.18em I}}
\def\IK{\relax{\rm I\kern-.18em K}}
\def\IL{\relax{\rm I\kern-.18em L}}
\def\IM{\relax{\rm I\kern-.18em M}}
\def\IN{\relax{\rm I\kern-.18em N}}
\def\IO{\relax\hbox{$\inbar\kern-.3em{\rm O}$}}
\def\IP{\relax{\rm I\kern-.18em P}}
\def\IQ{\relax\hbox{$\inbar\kern-.3em{\rm Q}$}}
\def\IR{\relax{\rm I\kern-.18em R}}
\font\cmss=cmss10 \font\cmsss=cmss10 at 7pt
\def\IZ{\relax\ifmmode\mathchoice
{\hbox{\cmss Z\kern-.4em Z}}{\hbox{\cmss Z\kern-.4em Z}}
{\lower.9pt\hbox{\cmsss Z\kern-.4em Z}}
{\lower1.2pt\hbox{\cmsss Z\kern-.4em Z}}\else{\cmss Z\kern-.4em Z}\fi}
\def\IGa{\relax\hbox{${\rm I}\kern-.18em\Gamma$}}
\def\IPi{\relax\hbox{${\rm I}\kern-.18em\Pi$}}
\def\ITh{\relax\hbox{$\inbar\kern-.3em\Theta$}}
\def\IOm{\relax\hbox{$\inbar\kern-3.00pt\Omega$}}
\def\CM {{\cal M}}

\def\CP {{\cal P }}
\def\CL {{\cal L}}

\def\CO {{\cal O}}

\def\CI  {{\cal I}}
\def\p {\partial}
\def\CS {{\cal S}}

\def\pb{\bar{\partial}}

\def\log {{\rm log}}

\def\c{\cdot}

\Title{ \vbox{\baselineskip12pt\hbox{hep-th/9310026}\hbox{YCTP-P19-93}}}
{\vbox{
\centerline{Symmetries}
\centerline{of the Bosonic String}
\centerline{S-Matrix} }}
\bigskip
\centerline{Gregory Moore}
\bigskip\centerline{moore@castalia.physics.yale.edu}
\smallskip\centerline{Dept.\ of Physics}
\centerline{Yale University}
\centerline{New Haven, CT \ 06511}
\bigskip
\bigskip

\noblackbox

\noindent
The bracket operation on mutually
local BRST classes may be combined with
Lorentz invariance and analyticity to write
an infinite set of finite difference relations on
string scattering amplitudes.  When combined
with some simple physical criteria these relations
uniquely determine the genus zero string
$S$-matrix for $N\leq 26$-particle
scattering in $\IR^{25,1}$ in terms of a single
parameter, $\kappa$, the string coupling.
We propose that the high-energy limit of the
relations are the Ward identities for the high-energy
symmetries of string theory.

\bigskip

\Date{October 8, 1993}
%\draft

\newsec{Introduction}

String theory is a generalization of gauge theory.
Historically, understanding the importance of
general covariance and
local isotopic-spin invariance were crucial
steps in the formulation
of general relativity and Yang-Mills theory.
One might therefore suspect that
a complete formulation of string theory
will be predicated upon a deeper understanding of
the symmetries in the theory.

In
\ref\grsslett{D. Gross, ``High energy symmetries of
string theory,'' Phys. Rev. Lett. {\bf 60B}(1988)1229}
D. Gross proposed that one useful tool for discovering
fundamental string symmetries is the analysis of
the high energy behavior of
string scattering amplitudes. In field theories of
spontaneously broken gauge invariance this technique
works very well. Using the saddle-point analysis of
\ref\gm{D.J. Gross and P.F. Mende, ``String theory
beyond the Planck scale,'' Nucl.Phys. B303: 407, 1988;
``The high-energy behavior of string scattering
amplitudes,'' Phys. Lett. 197B: 129, 1987.}\
Gross derived an infinite set of linear relations
between bosonic string 4-particle $S$-matrix amplitudes.
Usually linear relations between scattering amplitudes
are derived from an underlying symmetry, and it was
suggested in \grsslett\ that such a symmetry must
explain these linear relations.
Some work has subsequently been done with
a view towards understanding this mysterious symmetry in
\ref\wttlett{E. Witten, ``Spacetime and topological
orbifolds,'' Phys. Rev. Lett. {\bf 61B}(1988)670}
\ref\followup{M. Evans and B.A. Ovrut, ``Spontaneously
broken inter mass level symmetries in string theory,''
Phys. Lett. {\bf 231B}(1989)80;
``Deformations of conformal field theories and symmetries
of the string,'' Phys. Rev. {\bf D41}(1990)3149}
\ref\nullstring{
U. Lindstrom, B. Sundborg, and G.
Theodoridis, Phys. Lett. {\bf 258B}(1991)319;
``The zero tension limit of the spinning string,''
Phys. Lett. {\bf 258B}(1991)331}.

In
\ref\finite{G. Moore, ``Finite in all directions,''
hep-th/9305139; ``Symmetries and symmetry-breaking in
string theory,'' hep-th/9308052, to appear in Proceedings of
SUSY93.}
it was suggested that the infinite dimensional
hyperbolic symmetries which arise upon toroidal compactification
of time might be the source of the high energy symmetries
of string theory.
The original program of \finite\ is probably
misguided, for reasons explained in appendix B below.
Nevertheless,  as we show in the present paper,
the basic idea that generalized Kac-Moody
algebras are high-energy symmetries of string
theory is correct.  In fact, more is true. We can replace
high energy Ward identities, which relate amplitudes
at the same values of $s,t$ by finite difference
relations for the exact amplitudes. These finite
difference relations put strong constraints on
the genus zero string $S$-matrix. We will show that
they determine scattering amplitudes at all mass
levels in terms of tachyon scattering. Moreover,
when supplemented with some mild analyticity
requirements (e.g. Regge-like behavior at
$s\to \infty$)  the finite difference relations even
determine the tachyon amplitude itself. A technical
point discussed in section 3.3 limits our discussion
to $N$-particle scattering for $N\leq 26$.

In essence the answer to the problem posed in
\grsslett\ is very simple:
the underlying symmetries are the bracket algebras defined by
on-shell mutually local BRST invariant chiral vertex operators.

\newsec{Review of Bosonic String Scattering}

It is convenient to summarize some standard facts
\ref\grschwtt{M. Green,
J. Schwarz, and E. Witten, {\it Superstring theory\/},
Cambridge Univ. Press (1987).}.

\subsec{On-Shell States}

The states of the open bosonic string
are defined in terms of the BRST cohomology
$H^*$
\grschwtt
\ref\FGZ{I. Frenkel, H. Garland, and G. Zuckerman, ``Semi-infinite
cohomology and string theory,'' Proc. Nat. Acad. Sci. {\bf 83}(1986)8442}.
We focus on the chiral ghost number 1 cohomology,
$\CH=H^{g=1}$. The space $\CH$
 is graded by level number $n$ and momentum
$p\in \IR^{25,1}$:
\eqn\opcoh{
\CH=\oplus_{n\in \IZ^+} \int_{\IR^{25,1}} dp \quad \CH[p,n]
}
where $\CH[p,n]=0$ unless  $p^2=2-2n$ and
$\dim \CH[p,n]=p_{24}(n)$.
$\CH$ is an induced representation of the Poincar\'e group
in $\IR^{25,1}$.

Cohomology classes have representatives
of the
form  $cV$ where $V$ is a chiral vertex operator
satisfying the physical state conditions. These conditions
 state
that $V$ is a dimension one Virasoro primary.
\foot{The ghosts will not play an essential role in this
paper so we will often identify a class  with  $V$.}
Such operators have the form
$V= \CP e^{i p X}$ where $\CP$ is a polynomial
in $\p^* X$ of dimension $n$.
For example:
\eqn\states{\eqalign{
n=0:\qquad & \CP=1 \cr
n=1:\qquad & \CP=i \zeta\cdot \p X \qquad \zeta\in T\IR^{25,1}\cr
& \zeta\cdot p=0 \qquad \zeta\sim \zeta+ \lambda p\cr
n=2:\qquad & \CP=i p\c \zeta\cdot \p^2 X +
\p X \cdot  \zeta\cdot \p X\qquad \zeta\in
 \bigl(T\IR^{25,1}\bigr)^{\otimes 2}\cr
\tr(\zeta)-2 p\cdot  \zeta\cdot p=0& \qquad
 \zeta \sim  \zeta+\half\bigl[p\otimes \chi+\chi\otimes p-{1\over 3}(p\c
\chi)\eta\bigr] \cr}
}
The fields $X^\mu$ are
always normalized to have the correlator
$$\p X^\mu(z) \p X^\nu(w) \sim - {\eta^{\mu \nu}\over (z-w)^2 }\qquad .$$
We use units where $\alpha'=\half$ for open string
amplitudes.
Polarization tensors, or multiplets of polarization
tensors are generically denoted by $\zeta$.

\subsec{S-Matrix Amplitudes}

S-matrix amplitudes are multilinear functions
$\CA:\CH^{\otimes n}\to \IC$ constructed as
follows. The operator formalism associates a
measure $\Omega(V_1,\dots V_n)$ on the
moduli space $F$ of ordered points on the
boundary of the unit disk.
We define $\CA=\int_F \Omega$.
By Mobius invariance of $\Omega$, $\CA$ is invariant
under cyclic permutations.

The definition of $\int_F \Omega$ requires
some care. Using Mobius invariance of
$\Omega$  we
transform the disk to the upper half plane and
write:
\eqn\npti{
\CA(V_1,\dots V_n) = \kappa^{n-2}
\int_{F_{n-3}([0,1])} \prod_2^{n-2} dy_i
\langle 0|V_n(y_n) V_{n-1}(y_{n-1})\cdots V_1(y_1)|0\rangle
}
where $y_n=\infty, y_{n-1}=1,y_1=0$ and $F_{n-3}([0,1])$ is the
moduli space of $n-3$ ordered points on the interval.
The integral \npti\  is given meaning
as follows. The integrand is a function of the $y_i$ and
of the relativistic invariants $s_{ij}\equiv p_i\c p_j$.
It follows from the o.p.e. that there is
a domain where $Re(s_{ij})$
is sufficiently large and positive, or negative,
(depending on $ij$) so that the
integral is absolutely convergent and defines  a
holomorphic function of the $s_{ij}$. The analytic
continuation from this domain defines an
amplitude which is a meromorphic function of
the $s_{ij}$.

It will be important to specify clearly
the independent relativistic
invariants that $\CA$ depends upon. These
invariants are
formed out of momenta $p_i$ and polarization
tensors $\zeta_i$.  Relativistic invariants
formed from the $p_i$ alone parametrize
the different orbits of $SO(1,d-1)$ on the space of
$n-1$ independent momenta. Taking into consideration
 the relevant little group we see that
the number of independent
relativistic invariants for $n$ particle scattering
in $d$ dimensions is
\eqn\nbrinv{
\eqalign{
nd -\half d(d+1) & \qquad\qquad {\rm for}\quad  n\geq d+1\cr
\half n(n-1) & \qquad\qquad {\rm for}\quad  n\leq d+1\cr}
}
For $n\leq d$ a set of algebraically independent
invariants can be chosen to be an appropriate
collection of the $s_{ij}$. For example, we may
choose $s_{ij}$ for $1\leq i\leq j\leq n-1$.
(We make a different choice below.)
For $n\geq d+1$
the story is more complicated.
By classical invariant theory
\ref\weyl{H. Weyl, {\it The Classical Groups}, Princeton University
Press, Princeton 1946}
the ring of polynomial invariants is generated by
the $s_{ij}$ and by
$[p^{i_1},\dots, p^{i_d}]=
\epsilon^{\mu_1 \cdots \mu_d}
p^{i_1}_{\mu_1}\cdots p^{i_d}_{\mu_d}$.
The relations are generated by
$[p^{i_1},\dots, p^{i_d}][p^{j_1},\dots, p^{j_d}]=\det_{s,t} p^{i_s}\c
p^{j_t}$,
$\det \Delta=0$ where $\Delta$ is any $(d+1)$-dimensional
minor of the matrix $(s_{ij})$, and $\sum_{\sigma} \pm
[p^{i_{\sigma(1)}},\dots,p^{i_{\sigma(d)}}]p^{i_{\sigma(d+1)}}\c p^j=0$
\ref\weyli{See \weyl, pp. 75-78}.
Thus for $n\geq d+1$ a maximal algebraically
independent set of
invariants can be taken to be an appropriate set of
$s_{ij}$ together with $[p^{1},\dots, p^{d}]$.

We separate the independent relativistic
invariants into three types:

\noindent
$\bullet$ Levels of the particles:  $p_i^2=2-2n_i$, $i=1,\dots n$

\noindent
$\bullet$ Scalar product invariants. According
to the above discussion for $n\leq d$
these can be taken to
be the $s_{ij}$, $1\leq i<j\leq n-2$, together with
$1\leq i \leq n-3, j=n-1$ . For
$n\geq d+1$ we must choose an appropriate
set of $n(d-1)-\half d(d+1)-1$ $s_{ij}$'s.

\noindent
$\bullet$ Polarization invariants.  These are formed from contractions of
the $\zeta$'s with themselves or the $\zeta$'s with $p$'s.
The only way the relativistic invariants $[p^{i_1},\dots, p^{i_d}]$
can enter a (bosonic) string amplitude
is through the polarization invariants.

{\bf Example}:
The case $n= 4$ is of particular importance.
This is the
function $\CA:\CH^{\otimes 4}\to \IC$
given by:
\eqn\open{
\CA(V_1,V_2,V_3,V_4)=\kappa^2 \int_0^1 dz
\langle 0|V_4(\infty)V_3(1)
V_2(z)V_1(0)|0\rangle
}
The
kinematic invariants are $s=\ p_1\cdot p_2, t=p_2\cdot p_3$.
 We let
\eqn\indpinvt{
\biggl\{ \matrix {\zeta_1 & \zeta_2& \zeta_3 & \zeta_4\cr
p_1 & p_2 & p_3 &p_4 \cr}
\biggr\}
}
stand for an ordered set of independent polarization  invariants.
\foot{We choose a lexicographic ordering, with $\zeta$
in front of $p$.}
Thus,the amplitudes are functions with
independent arguments:
\eqn\arguments{
\CA_{n_1,n_2,n_3,n_4} \biggl(
\biggl\{ \matrix {\zeta_1 & \zeta_2& \zeta_3 & \zeta_4\cr
p_1 & p_2 & p_3 &p_4 \cr}
\biggr\} \biggl | s,\ t \biggr)
}

\subsec{Analyticity Properties}

As an analytic function $\CA$ is a polynomial in
the polarization invariants. The coefficients
of this polynomial are
meromorphic functions of the $s_{ij}$.
One of the goals of this paper is to
replace the complicated formula
\npti\ by a symmetry principle.
We will need to take the following
three  analyticity properties as axiomatic:

\noindent
{\bf AP1}: Location of poles.  The dual diagrams of the
genus zero open string amplitudes are binary trees rooted at
the center of the unit disk with ordered terminal vertices on the
boundary of the disk as in:

%%Begin InstantTeX Picture
\let\picnaturalsize=N
\def\picsize{2.5in}
\def\picfilename{tr.eps}
%If you do not have the picture file add:
%\let\nopictures=Y
%to the beginning of the file.
\ifx\nopictures Y\else{\ifx\epsfloaded Y\else\input epsf \fi
\let\epsfloaded=Y
\centerline{\ifx\picnaturalsize N\epsfxsize \picsize\fi
\epsfbox{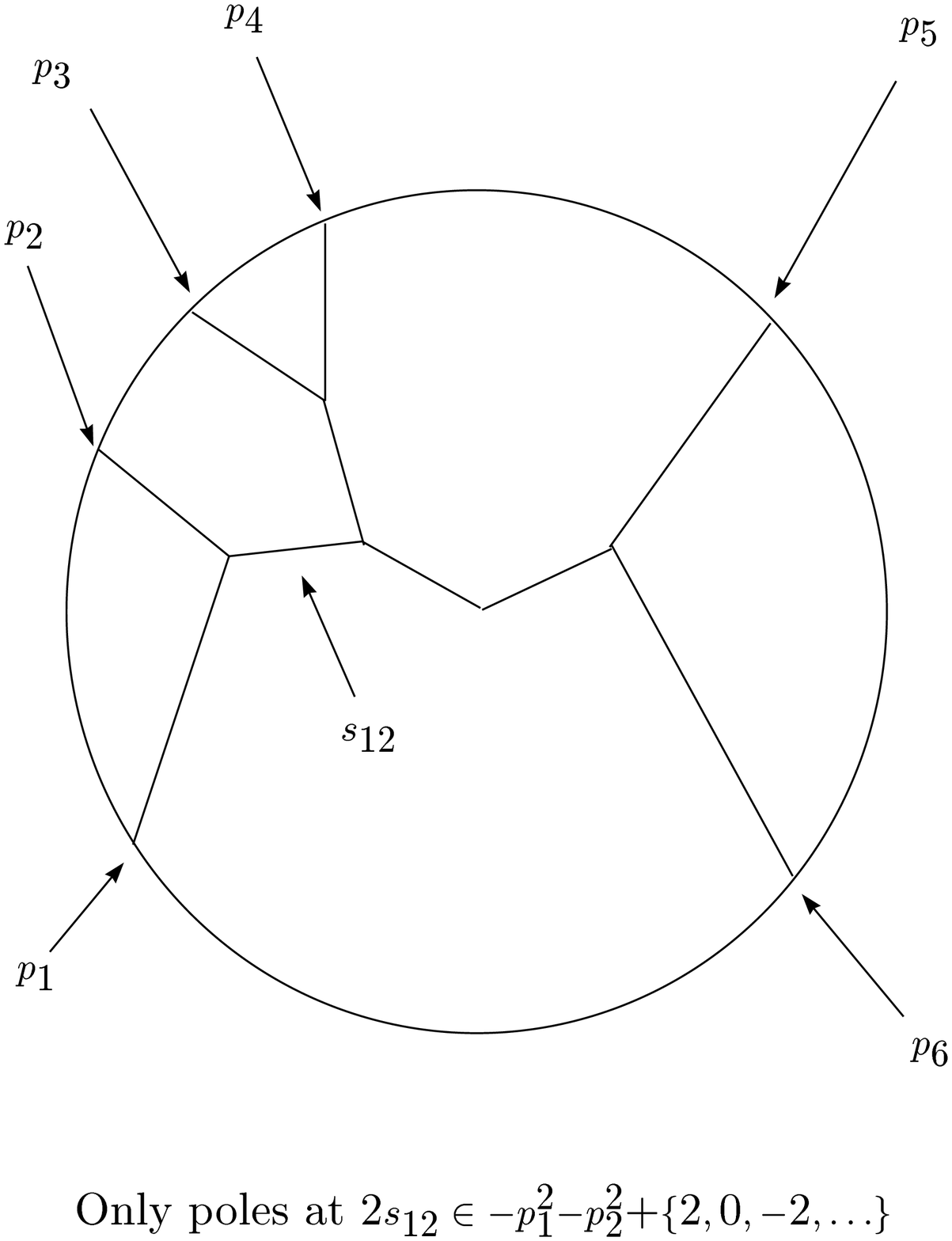}}}\fi
%%End InstantTeX Picture

\bigskip
Each terminal vertex carries an ingoing momentum
$p_i$. Using momentum conservation we associate
a momentum $p_I$ with each internal edge $I$.
We assume that the  amplitudes can only have poles when
$p_I^2\in \{ 2,0,-2,-4,\dots \}$ for some $I$ in some
dual diagram. Explicitly, this is the condition that
$(p_i+\cdots + p_{i+k})^2 \in \{ 2,0,-2,-4,\dots \}$
for some $i,k$, where indices are understood modulo $n$.

The proof  that the $n$-point amplitudes \npti\
satisfy the axiom {\bf AP1} makes use of the
operator product expansion and Mobius invariance
of $\Omega$.

\noindent
{\bf AP2}: Growth at infinity. We assume that
the amplitude has at most power law growth as
any $s_{ij}$ tends to infinity, holding all other
independent kinematic invariants fixed. This is a
generalization of Regge behavior.

The axiom {\bf AP2} can be motivated by
considering the $n$-tachyon scattering
amplitude:
\eqn\kbnielsen{
\CA=\kappa^{n-2}
\int_0^1 \prod_2 ^{n-2} dy_i\theta(y_i-y_{i-1})
\prod_{1\leq i<j\leq n-1} (y_j-y_i)^{s_{ij}}
}
where $y_{n-1}=1, y_1=0$.
 The standard argument
\ref\mandelstam{S. Mandelstam, ``Dual Resonance
Models,'' Phys. Reports, {\bf 13} (1974)259}
proceeds as follows. If $Re(s_{ij})\to +\infty$ then since all
the factors in the product in \kbnielsen\ are $\leq 1$
the integration is dominated by the region where
those factors raised to a power $s_{ij}$ are near one.
Making an exponential change of variables and
isolating this region proves the claim.  As the simplest
example of this argument, suppose we take the $s_{ij}$
appearing in \kbnielsen\  as independent variables
($s_{1,n-1}$ will be considered dependent). Consider
the limit $Re(s_{12})\to +\infty$ holding all other independent
$s_{ij}$ fixed.
We write $y_2=e^{-x_2}$ and the
integral is dominated by the region $x_2\sim 0$.
Making this approximation in the rest of the integrand
we find the asymptotics $s_{12}^{-s_{2,n-1}}$. In
general we find
$$\CA\sim \alpha s_{ij}^\beta$$
where
$\beta$ is a combination of the other $s_{kl}$'s.
This argument
extends to any correlation function and justifies the
axiom {\bf AP2}.

Finally we need an axiom that relates the amplitudes
for different values of $n$. These are the well-known
and standard tree-level unitarity equations:

\noindent
{\bf AP3}:  Factorization. If we cut a dual diagram
on some internal edge $I$ we decompose it into
two dual diagrams $I_1,I_2$.  We assume that
when $p_I^2\to 2-2n_I$ and other momenta are
in general position the residue at the pole is
\eqn\respole{
\sum_{a,b} \CA(V_i,\dots, V_{i+k},V_a)
G^{ab} \CA(V_b, V_{i+k+1},\dots,
V_{i-1})
}
where $\{ V_a\}$ is a basis for $\CH[p_I,n_I]$ and
$G^{ab}$ is the inverse of the positive definite metric
on $\CH[p_I,n_I]$ whose existence is assured by the
no-ghost theorem.

The proof that property {\bf AP3} is satisfied again
makes use of the operator product expansion.

{\bf Remark}: In stating the factorization axiom we
have not made any use of a complex structure or
of local coordinates on the disk. One could state
a stronger factorization axiom valid for all values of
the momenta. This
requires the introduction of the
full off-shell BRST chain complex and the introduction
of local coordinates. We will not need that in the
present paper.

\subsec{High Energy-Fixed Angle Scattering}

We define
$s=p_1\cdot p_2 \equiv -2 E^2$, and
$t=p_2\cdot p_3 \equiv 2 E^2 \sin^2\half \theta$.
In the limit of high energy scattering, where all masses
are effectively zero, $E$ and $\theta$ have the physical interpretation
of center of mass energy and scattering angle. In the plane
of scattering the spatial momenta look like:

%%Begin InstantTeX Picture
\let\picnaturalsize=N
\def\picsize{2.5in}
\def\picfilename{hs.eps}
%If you do not have the picture file add:
%\let\nopictures=Y
%to the beginning of the file.
\ifx\nopictures Y\else{\ifx\epsfloaded Y\else\input epsf \fi
\let\epsfloaded=Y
\centerline{\ifx\picnaturalsize N\epsfxsize \picsize\fi
\epsfbox{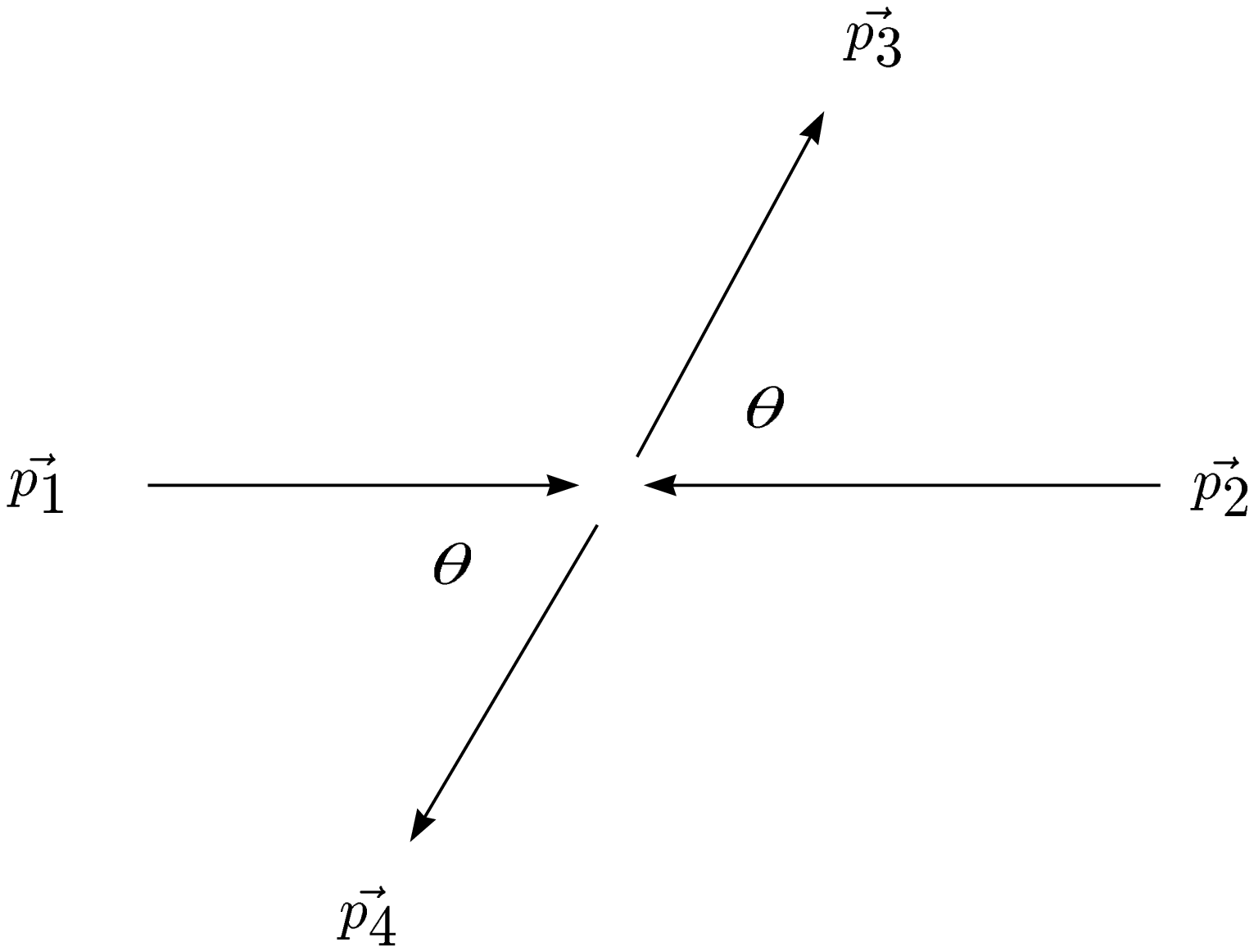}}}\fi
%%End InstantTeX Picture

The high
energy limit is defined to be the limit where
$E^2\to \infty$ along any ray other than
the positive or negative real axis, holding $\theta$ and all other
independent relativistic invariants fixed.
\foot{Warning: this differs slightly from the
$\alpha'\to \infty$ limit of \grsslett, since we hold
expressions of the form $\zeta \cdot p^k$ constant, even
for polarizations with longitudinal components.}

Gross and Mende studied the asymptotic behavior of the
amplitudes in the high-energy limit \gm.
The amplitudes are dominated by a saddle point. For open
strings:
\foot{The result is BRST invariant despite
appearances. The choice of the point $z_0$ magically cancels
all differences arising from different choices of representative
 $\CP$ for the cohomology class.}
\eqn\gmthm{\eqalign{
\CA &\sim \CA^{s.p.} \biggl[1+\CO\bigl(1/s,1/t,1/(s+t)\bigr)\biggr]\cr
\CA^{s.p.} &\equiv\sqrt{2 \pi}\sqrt{st\over (s+t)^3}
\langle 0|V_4(\infty)V_3(1)V_2(z_0)V_1(0)|0\rangle\cr}
}
where $z_0\equiv s/(s+t)=1/\cos^2\half \theta$.

The proof of \gmthm\ may be obtained by combining
Stirling's formula with the analytic structure of the
amplitudes described above. Alternatively, one may
use  a saddle point
analysis of either the integral over moduli space or of
the path integral. The saddle
point typically lies outside the domain of integration,
and for good reason. The true high energy asymptotics
obtained by taking $E^2\to \infty$ along the positive
real axis must encounter an
infinite set of poles when any intermediate state
goes onshell. These poles are manifestly absent from
\gmthm. The saddle point analysis applies to the asymptotics
of analytically continued $S$-matrix elements. This is
important to bear in mind when considering generalizations
to $n\geq 5$ point functions and to higher orders of
string perturbation theory, in which case one
probably must consider
asymptotics only for $E^2\to \pm i \infty$.

\newsec{Bracket Relations}

\subsec{Bracket}

Two years ago it was shown that the BRST cohomology
of 2D string theory has a rich algebraic
structure related to BV quantization
\ref\grndrng{E. Witten,
Ground ring of two-dimensional string theory'' (hep-th/9108004),
Nucl. Phys. B373 (1992) 187.}
\ref\witzwie{E. Witten and
B. Zwiebach, ``Algebraic Structures and
Differential Geometry in
2D String Theory'' (hep-th/9201056),
Nucl. Phys. B377 (1992) 55.}
\ref\vermster{E. Verlinde, ``The master equation of
2D string theory,'' hep-th/9202021; Nucl. Phys. B381 (1992) 141.}.
Subsequently,  Lian and Zuckerman
studied
the algebraic structures of the BRST
cohomology based on an arbitrary
chiral algebra (=vertex operator algebra
\ref\flm{I.B. Frenkel, J. Lepowsky, and A. Meurman,
``Vertex Operator Algebras and the Monster,''
Academic Press, New York, 1988}) in
 \ref\lziii{B. Lian and G. Zuckerman,
``New perspectives on the brst algebraic structure of string theory,''
(hep-th/9211072) Commun. Math. Phys. {\bf 154}613 (1993).}
\ref\relres{Related results have been obtained in
G. Segal, Lectures at the Isaac Newton Institute,
August 1992, and lectures at Yale University, March 1993;
E. Getzler, ``Batalin-Vilkovisky Algebras and
Two-Dimensional Toplogical Field Theories'' (hep-th/9212043)}.
As shown by Lian and Zuckerman, there is
an operation $\{ \c , \c \}: H^{g_1}\times H^{g_2}\to
H^{g_1+g_2-1}$ which they called the
``Gerstenhaber bracket.''  The bracket is very
fundamental: it exists for  arbitrary bosonic string
cohomology and plays a role related to the
BV anti-bracket for
the on-shell BV structure of string theory.
In the present case we may identify ghost number 1
classes
with chiral physical state operators and
the bracket is simply the standard
``commutator'' of dimension one currents:
\eqn\brckt{
J\otimes V \to \{J,V\}(z) \equiv \oint_z dw J(w) V(z)
}
In general, the ghost number one cohomology
based on a chiral algebra is a Lie algebra.

In our case $\CH$ in \opcoh\ is {\it not} based on a
chiral algebra because the fields have monodromy
when considered as chiral vertex operators.
Nevertheless, when $J,V$ are mutually local
\brckt\  still  makes sense.
\foot{In this paper ``mutually local'' means the o.p.e.
has only integral powers of $z-w$. Equivalently, the
monodromy is trivial (although the braiding matrix
could be $(-1)$).} In particular,
given two momenta $p$, $q$ with
$q^2=2-2n_1$, $p^2=2-2n_2$, $(p+q)^2=2-2n_3$.
we have a map:
$$\{ \c , \c \}:\CH[q,n_1]\otimes \CH[p,n_2] \to \CH[p+q,n_3]$$
We may extend the bracket to all of $\CH$:
$$\{ \c , \c \}:\CH\otimes \CH\to \CH$$
by defining it to be  zero on pairs not mutually local.
\foot{Alternatively, one can take the point of view
that the bracket is only defined on {\it some} pairs of
states. That is, it is like multiplication in a groupoid,
or composition of morphisms in a category.}

The bracket has the two properties:

{\bf B1}. If $p_1,p_2$ are the momenta of $V_1,V_2$ then:
$$ \{ V_{1},V_{2} \}=-(-1)^{p_1\c p_2} \{ V_{2}, V_{1} \}$$

{\bf B2}. If $V_1,V_2,V_3$ are all pairwise mutually local then:
\eqn\grdjc{
(-1)^{p_1\c p_3} \{ V_1, \{ V_2,V_3\} \}
+(-1)^{p_3\c p_2} \{ V_3, \{ V_1,V_2\} \}
+(-1)^{p_2\c p_1} \{ V_2, \{ V_3,V_1\} \} = 0
}

Unfortunately \grdjc\ does not hold for all triples in
$\CH$.

The bracket on ghost number one cohomology
generalizes the Lie algebra of the Gerstenhaber
bracket  in two ways. First,  property {\bf B1} shows
that the bracket is
``vectorially-graded.''  Second, the
vectorially-graded Jacobi relation only holds for mutually
local triples.

A table of useful structure constants for $\{ \c ,\c \}$
can be found in appendix A.

\subsec{Relations}

Let $J$ be any chiral physical state operator
of momentum $q$ and let
$V_i$, $i=1,2,3,4$ be  chiral
physical state operators of momenta $p_i$
such that $q+\sum p_i=0$.
Assume that $q\cdot p_i$ are integral so that $J$ and $V_i$
are mutually local.
\foot{ We do
{\it not} assume that the $V_i$ are mutually local.}
In this case we can regard  the integrand of \open\
as a correlator of chiral vertex operators for conformal
field theory on the plane.
Using standard contour deformation  arguments
we derive  the identity
\eqn\cftwi{\eqalign{
0 &= \langle 0|V_4 (\infty)V_3(1) V_2(z)\{J, V_1\}(0)|0\rangle\cr
+& (-1)^{q\cdot p_2}\langle 0|V_4(\infty)V_3(1)
\{J, V_2\}(z) V_1(0)|0\rangle\cr
+& (-1)^{q\cdot p_2+q\cdot p_3}\langle 0|V_4(\infty)\{J,V_3\}(1)
 V_2(z)V_1(0)|0\rangle\cr
&+ (-1)^{q\cdot p_2+q\cdot p_3+q\cdot p_4}
\langle 0|\{J,V_4\}(\infty)V_3(1) V_2(z)V_1(0)|0\rangle\cr}
}

Now we simply integrate $z$ from $0$ to $1$, assuming
that $Re(s)$ and $Re(t)$ are sufficiently positive that
the integral is absolutely convergent.  This gives the
finite difference relations:
\eqn\fdr{\eqalign{
0=& \CA_{\tilde n_1,n_2,n_3,n_4} \biggl(
\biggl\{ \matrix {\tilde \zeta_1 & \zeta_2& \zeta_3 & \zeta_4\cr
p_1 +q& p_2 & p_3 &p_4 \cr}
\biggr\} \biggl | s+q\c p_2,\ t \biggr)
 \cr
+& (-1)^{q\cdot p_2}\CA_{n_1,\tilde n_2,n_3,n_4} \biggl(
\biggl\{ \matrix {\zeta_1 & \tilde \zeta_2& \zeta_3 & \zeta_4\cr
p_1 & p_2+q & p_3 &p_4  \cr}
\biggr\} \biggl | s+q\c p_1,\ t +q\c p_3\biggr)
\cr
+& (-1)^{q\cdot p_2+q\cdot p_3}\CA_{n_1,n_2,\tilde n_3,n_4} \biggl(
\biggl\{ \matrix {\zeta_1 & \zeta_2& \tilde \zeta_3 & \zeta_4\cr
p_1 & p_2 & p_3+q & p_4 \cr}
\biggr\} \biggl | s,\ t + q\c p_2\biggr)
\cr
+& (-1)^{q\cdot p_2+q\cdot p_3+q\cdot p_4}\CA_{n_1,n_2,n_3,\tilde n_4}
 \biggl(
\biggl\{ \matrix {\zeta_1 & \zeta_2& \zeta_3 & \tilde \zeta_4\cr
p_1 & p_2 & p_3 &p_4 +q \cr}
\biggr\} \biggl | s,\ t \biggr)
\cr}
}
Here $\tilde n$, $\tilde \zeta$ refer to the transformed operator
under the bracket.

The relations \fdr\  are  finite difference equations relating
scattering amplitudes for particles at different mass levels.
The essential content of these identities  is
displayed by a matrix of the levels of the states
involved. Each row of the matrix encodes the
level-number of the states in an amplitude.
The general Ward identity is  associated with a
matrix of the form:
\eqn\levelmatrix{
\pmatrix{
\tilde n_1 & n_2 & n_3 & n_4\cr
n_1 &\tilde n_2 & n_3 & n_4\cr
n_1 & n_2 & \tilde n_3 & n_4\cr
n_1 & n_2 & n_3 &  \tilde n_4\cr}
}
Here  integers $n_i\geq 0$, are levels of the untransformed
particles, while the integers $\tilde n_i$ are the levels of the
transformed particles. The level $n_J$ of the symmetry
current $J$ is easily extracted from the
sum of the diagonal minus the anti-diagonal:
\eqn\sumrule{
\sum \tilde n_i-\sum n_i=-q^2=2 n_J-2}
If $\tilde n_i$ is negative then $\{J,V\}=0$ and
the corresponding amplitude simply vanishes.

This procedure generalizes
to $N$-particle scattering.  Whenever
$J$ of momentum $q$
is mutually local w.r.t all the $p_i$
and $q+\sum p_i=0$  we have an identity
\eqn\genfdr{
\sum_i (-1)^{q\c p_2+\cdots q\c p_i}
\CA\bigl(V_1,\dots, \{ J, V_i \} ,\dots V_N\bigr) =0
}
These relations are subject to
an important technical restriction discussed in the
next section.

{\bf Remark}: Although we have used
currents $J$ to derive relations
on amplitudes, they are not unbroken symmetry currents
in the theory since there are many operators in the
BRST cohomology which are {\it not} mutually local
w.r.t. any given $J$.  Nevertheless, using
Lorentz invariance and analyticity we are still
able to derive relations on amplitudes.

\subsec{Existence of required momenta}

The identities \fdr\ would be of little use if
one could rarely choose momenta in the
way we have indicated.  In this section we show
that we can always find momenta corresponding
to any prescription of the levels and kinematic invariants
(at least, for $N\leq d=26$).
In doing so it is necessary to allow the momenta to be
complex.
We may do this since the BRST
conditions and the vertex operator calculus
make perfect sense when momenta are complex.

\bigskip
\noindent
{\bf Lemma}.  Suppose $N\leq d$.
For any complex numbers
$n_i$, $\tilde n_i$, $i=1,\dots N$, and
$\half N(N-3)$ complex numbers
$z_{ij}, 1\leq i < j\leq N-2$, $z_{1,N-1},\dots z_{N-3,N-1}$,
there exist
momenta
\foot{The momenta are allowed to take
values in the extended complex plane $\hat{\IC}$.}
$p_1,\dots, p_N\in \hat{\IC}^d$
such that, if we define $q\equiv -\sum p_i$ then
\eqna\conds
$$\eqalignno{p_i^2  &=2-2n_i &\conds a\cr
(p_i+q)^2 =\bigl(\sum_{j:j\not= i} p_j\bigr)^2&=2-2\tilde n_i
&\conds b\cr
p_i\c p_j &= z_{ij}&\conds c\cr}$$

Proof: For $N\leq d$ the invariants $s_{ij}$ made from
$N$ momenta $p_i$ are algebraically independent.
It is straightforward to solve  \conds\  as a linear
system of equations to find the $p_i\c p_j$ in terms
of linear combinations of the $n_i,\tilde n_i, z_{ij}$.
We can  regard the  equations for $p_i\c p_j$ as equations
for the intersection of $\half N(N+1)$ quadrics in $\IP^{Nd}$.
These intersect in a variety of codimension at
most $\half N(N+1)$
\ref\kendig{See, e.g., K. Kendig, {\it Elementary
Algebraic Geometry}, Springer, 1977, thm. 3.8}. Even if
this variety lies at infinity we can use the solution - we
simply must take a momentum $\to \infty$ limit in the
amplitudes. $\spadesuit$

The different sets of polarization invariants in \fdr\
are polynomially related.
Therefore, using the analyticity of $\CA$ we conclude that
\fdr\ holds for all values of $s,t$ and all polarization tensors
satisfying these polynomial relations.
Similarly,
provided $n_i, n_J$ are positive integers
and $\tilde n_i$ are integers, we can
find momenta for
arbitrary kinematic invariants for which the identities
\genfdr\ can be written.

Unfortunately one cannot generalize the
lemma to the case $N>d$.
For $N\geq d+1$ the
$s_{ij}$ are not algebraically indpendent, hence
we cannot specify arbitrarily
the $n_i, \tilde n_i$ and independent
$s_{ij}, i\not=j$ (for $N$-particle scattering).

We are therefore
stuck with the rather distasteful limitation to
$N$-particle scattering for $N\leq d$.

\subsec{High Energy Limit}

Combining  \gmthm\   with \fdr\  we obtain the
high energy identities:
\eqn\wi{\eqalign{
0=& z_0^{p_2\cdot q}\CA^{s.p.}(\{J, V_1\},
V_{2},V_{3},V_{4})\cr
+& (-1)^{q\cdot p_2}z_0^{p_1\cdot q}(1-z_0)^{p_3\cdot q}\CA^{s.p.}
( V_{1},\{J, V_2\},V_{3},V_{4})\cr
+& (-1)^{q\cdot p_2+q\cdot p_3}(1-z_0)^{p_2\cdot q}\CA^{s.p.}
( V_{1},V_{2},\{J,V_3\},V_{4})\cr
+& (-1)^{q\cdot p_1}\CA^{s.p.}
( V_{1},V_{2}, V_{3},\{J,V_4\})\cr}
}
where $z_0=s/(s+t)=1/\cos^2\half \theta$,
$1-z_0=t/(s+t)=-\tan^2 \half \theta$, and
each amplitude in \wi\ is evaluated at the {\it same}
value of $s,t$.  We take $s,t \to \infty$ holding
$q\c p_i$ fixed.
The amplitude $\CA^{s.p.}$
in \gmthm\ has the form $R U$ where $R$ is a rational
function of $s,t$ and
\eqn\gmfactor{
\eqalign{
U &\equiv
\sqrt{s t\over (s+t)^3} exp \biggl\{\bigl[ t\log t +s\log|s|-(s+t)\log|s+t|
\bigr]\biggr\} \cr
&={\cos\half\theta\over \sqrt{2} E \sin^3 \half\theta}
exp \biggl\{2E^2 \bigl(\sin^2 \half\theta\log[\sin^2\half\theta]
+\cos^2 \half\theta\log[\cos^2\half\theta] \bigr)\biggr\}\cr}
}
Shifts such as $s,t\to s+q\c p_2, t+q\c p_3$ lead to an
order one change in the amplitude from the exponential factor
in \gmfactor, leading
to the ``extra''
powers of $z_0,1-z_0$ in \wi. The change $\delta z_0$
in the position of the saddle-point in moduli space is
$\CO\bigl(1/s,1/t,1/(s+t)\bigr)$. Similarly, the shifts
in $s,t$ change
the rational function $R$ by terms of the same order. Hence
\wi\ holds up to factors $1+\CO\bigl(1/s,1/t,1/(s+t)\bigr)$.

\newsec{Six  Examples}

{\bf Example 1}:  The simplest example of \fdr\
has  level matrix:
\eqn\moi{
\pmatrix{0&0&0&0\cr
0&0&0&0\cr
0&0&0&0\cr
0&0&0&-2\cr}
}
{}From the entries we read off that this is a ``tachyonic
identity,''  i.e. $q^2=2$. We also can read off:
$$p_i^2=2\qquad \qquad p_1\c q=p_2\c q=p_3\c q=-1\qquad
\qquad p_4\c q=+1$$
Using  (A.1) of  appendix A we see
that \fdr\  implies
\eqn\fdri{
\CA(s-1,t)+\CA(s,t-1)=\CA(s-1,t-1)
}
where $\CA$ is the basic  Veneziano amplitude
$\CA_{0000}=\CA(s,t)$
for scattering
of four states at level $0$.

{\bf Example 2}: Consider the level matrix:
\eqn\mi{
\pmatrix{1&0&0&0\cr
0&0&0&0\cr
0&0&0&0\cr
0&0&0&-1\cr}
}
This encodes a lightlike Ward identity with $J=i \zeta_1\c \p X e^{i q\c X}$
and $q^2=0$. It relates $\gamma TTT$ scattering to $TTTT$
scattering, where $\gamma$ refers to the level one
photon and $T$ refers to the level zero tachyon.
For the  $\gamma TTT$ amplitude \arguments\
simplifies to:
$$\CA_{1000}(\zeta_1\cdot p_2,\zeta_1\cdot p_3|s,t)$$
{}From \mi\ we read off
$$\eqalign{
p_1^2=p_2^2=&p_3^2=p_4^2=2\cr
p_1\c q=-1\quad p_2\c q=p_3\c q&=0 \quad p_4\c q=+1\cr}
$$
Using $(A.3)$ we have:
\eqn\fdrii{
\CA_{1000}(\zeta_1\cdot p_2,\zeta_1\cdot p_3|s,t)=
- \zeta_1\cdot p_2\CA(s-1,t)
- \zeta_1\cdot p_3\CA(s,t)
}
{}From \wi\  we see that at high
energies $\gamma TTT$ scattering is related to
tachyon scattering via
\eqn\reli{
0=\CA_{1000}^{s.p.}(\zeta_1\cdot p_2,\zeta_1\cdot p_3|s,t)
+(1+t/s) \zeta_1\cdot p_2\CA^{s.p.}(s,t)
+ \zeta_1\cdot p_3\CA^{s.p.}(s,t)
}

Similarly, one easily derives relations on
similar amplitudes. For example:
\eqn\fdrvii{
\CA_{0100}(\zeta_2\cdot p_1,\zeta_2\cdot p_3|s,t)=
\zeta_2\cdot p_1\CA(s-1,t)
- \zeta_2\cdot p_3\CA(s,t-1)
}

{\bf Example 3}:  We now consider the level matrix
\eqn\mii{
\pmatrix{1&1&0&0\cr
0&1&0&0\cr
0&1&0&0\cr
0&1&0&-1\cr}
}
again giving a lightlike Ward identity. This relates
$\gamma \gamma TT$ to $\gamma TTT$ scattering.

We consider the $\gamma \gamma TT$ amplitude to be a function
of seven arguments:
$$\CA_{1100}(\zeta_1\cdot \zeta_2,\zeta_1\cdot p_2,
\zeta_1\cdot p_3,\zeta_2\cdot p_1,\zeta_2\cdot p_3|s,t)$$
Using  $(A.3)$  and  $(A.5)$ \fdr\ becomes
(after
shifting arguments and setting
$\zeta_2 \cdot q=0$):
\eqn\fdriii{
 \eqalign{
\CA_{1100}(x_1,x_2,x_3,x_4,x_5|s,t)&=
-\CA_{0100}(x_2 x_4-x_1,x_2 x_5|s-1,t) \cr
&\quad -x_3 \CA_{0100}(x_4, x_5,s,t)\cr
=(x_1-x_2 x_4)\CA(s-2,t)+ & x_2 x_5   \CA(s-1,t-1)
-x_3 x_4 \CA(s-1,t)\cr
&+x_3 x_5 \CA(s,t-1)\cr}
}
Which has the high energy limit:
\eqn\relii{\eqalign{
\CA_{1100}^{s.p.}(x_1,x_2,x_3,x_4,x_5|s,t)&=
-(1+t/s)\CA_{0100}^{s.p.}(x_2 x_4-x_1,x_2 x_5|s,t)\cr
&\qquad\quad -x_3 \CA_{0100}^{s.p.}(x_4,x_5|s,t)\cr}
}

Notice that if we were restricted to using real momenta and
polarizations then $p^2=q^2=p\cdot q=0 \Rightarrow p \parallel q$
and we would not derive relations on the most
general amplitude (e.g. $s$ would be fixed).
 This is not true if we
use complex momenta. After we derive the Ward identity
we can specialize to physical values of the relativistic
invariants, and, as in \relii, solve for the amplitude in
terms of previously known amplitudes.

{\bf Example 4}:
We now consider
\eqn\miv{
\pmatrix{1&1&0&0\cr
1&1&0&0\cr
1&1&1&0\cr
1&1&0&-1\cr}
}
which will express $\CA_{1110}$ in terms of
amplitudes at lower total level.
The ordered set of relativistic invariants
in $\CA_{1110}$ is:
\eqn\ordrd{
\biggl\{ \matrix {\zeta_1 & \zeta_2& \zeta_3 & 1\cr
p_1 & p_2 & p_3 & p_4 \cr}
\biggr\} =
\{\zeta_1\c \zeta_2, \zeta_1\c \zeta_3,
\zeta_2\c \zeta_3,\zeta_1\c p_2,
\zeta_1\c p_3,\zeta_2\c p_1,\zeta_2\c p_3,
\zeta_3\c p_1,\zeta_3\c p_2 \}
}

Following the procedure of the previous two examples we get:
\eqn\fdriv{\eqalign{
\CA_{1110}(x_1,\dots ,x_9|s,t)&=
\CA_{1100}(x_1 x_8,\, x_4 x_8,\, x_5 x_8-x_2,\, x_6,\, x_7| s,t)\cr
&\qquad +
\CA_{1100}
(x_1 x_9,\, x_4,\, x_5,\, x_6 x_9,\, x_7 x_9-x_3| s,t-1)\cr}
}
which has the high energy limit:
\eqn\reliv{\eqalign{
\CA_{1110}^{s.p.}(x_1,\dots ,x_9|s,t)&=
\CA_{1100}^{s.p.}(x_1 x_8,\, x_4 x_8,\, x_5 x_8-x_2,\, x_6,\, x_7| s,t)\cr
&\qquad +(1+s/t)
\CA_{1100}^{s.p.}
(x_1 x_9,\, x_4,\, x_5,\, x_6 x_9,\, x_7 x_9-x_3| s,t)\cr}
}

{\bf Example 5}:  We take
\eqn\mv{
\pmatrix{1&1&1&0\cr
1&1&1&0\cr
1&1&0&0\cr
1&1&1&1\cr}
}
which will relate the 4-photon amplitude to the
tachyon amplitude. We now need the bracket $(A.4)$

The ordered set of relativistic invariants
occurring in $\CA_{1111}$ is:
\eqn\ordrdi{\eqalign{
\biggl\{ \matrix {\zeta_1 & \zeta_2& \zeta_3 & \zeta_4\cr
p_1 & p_2 & p_3 & p_4 \cr}
\biggr\} =
\{\zeta_1\c \zeta_2, &\zeta_1\c \zeta_3,\zeta_1\c \zeta_4,
\zeta_2\c \zeta_3,\zeta_2\c \zeta_4,\zeta_3\c \zeta_4,\cr
\zeta_1\c p_2,\zeta_1\c  & p_3,\zeta_2\c p_1,\zeta_2\c p_3,
\zeta_3\c p_1,\zeta_3\c p_2,\zeta_4\c p_1,\zeta_4\c p_2 \} \cr}
}
So the four-photon amplitude is a function of $16$ arguments,
and \fdr\  becomes
\eqn\fdrv{\eqalign{
\CA_{1111}(x_1,\dots,&,x_{14}|s,t)\qquad\qquad\cr
=-\CA_{1110}&(x_1 x_{13},x_2 x_{13},x_4,x_7 x_{13},
x_8 x_{13}+x_3,x_9,x_{10},x_{11},x_{12}|s,t)\cr
-
\CA_{1110}&
(x_1 x_{14},x_2 ,x_4 x_{14},x_7,x_{8},
x_9 x_{14},x_{10} x_{14}+x_5, x_{11}, x_{12}|s,t+1)\cr
+x_6 \CA_{1100}&(x_1 ,x_7 ,
x_8 ,x_9,x_{10}|s,t)\cr}
}
with high energy limit:
\eqn\relv{\eqalign{
\CA_{1111}^{s.p.}(x_1,\dots,&,x_{14}|s,t)\qquad\qquad\cr
=-\CA_{1110}^{s.p.}&(x_1 x_{13},x_2 x_{13},x_4,x_7 x_{13},
x_8 x_{13}+x_3,x_9,x_{10},x_{11},x_{12}|s,t)\cr
-{t\over (s+t)}
\CA_{1110}^{s.p.}&
(x_1 x_{14},x_2 ,x_4 x_{14},x_7,x_{8},
x_9 x_{14},x_{10} x_{14}+x_5, x_{11}, x_{12}|s,t)\cr
+x_6 \CA_{1100}^{s.p.}&(x_1 ,x_7 ,
x_8 ,x_9,x_{10}|s,t)\cr}
}

{\bf Example 6}: As a final example we
look at a timelike identity with level matrix:
\eqn\mvi{
\pmatrix{2&0&0&0\cr
0&0&0&0\cr
0&0&0&0\cr
0&0&0&0\cr}
}
Hence $q^2=-2$ and
\eqn\tmj{\eqalign{
J=V_{\zeta,q}&\equiv \bigl[iq \c \zeta\c\p^2 X + \p X\c \zeta\c\p X\bigr]
e^{i q\c X}\cr
\tr( \zeta)-&2 q\cdot \zeta\cdot q=0\cr}
}
We parametrize the scattering of 1 level 2 on 3 level 0
states by the function of 8 variables:
\eqn\tTTT{
\CA_{2000}(p_1\c \zeta\c p_1,p_1\c \zeta\c p_2,
p_1\c \zeta\c p_3,p_2\c \zeta\c p_2,p_2\c \zeta\c p_3, p_3\c \zeta\c p_3|s,t)
}
Now we use the bracket in $(A.6)$.
After shifting arguments a little one finds the bracket
relation:
\eqn\fdrvi{\eqalign{
\CA_{2000}(p_1\c \zeta\c p_1,\dots p_3\c \zeta\c p_3|s,t)&
=-(p_2\c \zeta\c p_2+ p_1\c \zeta\c p_2)\CA(s-2,t+1)\cr
&+(p_3\c \zeta\c p_3+ p_1\c \zeta\c p_3)
\CA(s-1,t+1)\cr
-(p_3\c \zeta\c p_3+ p_1\c \zeta\c p_3+&
p_2\c \zeta\c p_2+ p_1\c \zeta\c p_2+2 p_2\c \zeta\c p_3)
 \CA(s-1,t)\cr}
}
with high energy limit:
\eqn\relvi{\eqalign{
\CA_{2000}^{s.p.}(p_1\c \zeta\c p_1,\dots p_3\c \zeta\c p_3|s,t)&
=-{t\over s}(1+{t\over s})(p_2\c \zeta\c p_2+ p_1\c \zeta\c
p_2)\CA^{s.p.}(s,t)\cr
&+{t\over s}(p_3\c \zeta\c p_3+ p_1\c \zeta\c p_3)
\CA^{s.p.}(s,t)\cr
-(1+{t\over s})(p_3\c \zeta\c p_3+ p_1\c \zeta\c p_3+&
p_2\c \zeta\c p_2+ p_1\c \zeta\c p_2+2 p_2\c \zeta\c p_3)
 \CA^{s.p.}(s,t)\cr}
}

As an exercise the reader may care to work out some
futuristic identities for $\CA_{2001}$.

\newsec{Determination of Tachyon Amplitudes}

In the previous section we saw that the relations
\fdr\ lead to a host of interlevel amplitude identities.
In the present section we discuss finite difference
relations on the tachyon amplitudes themselves.
We show that  relations \fdr\genfdr\  together with the
analyticity properties {\bf AP1},{\bf AP2} determine
the $N$- tachyon scattering
amplitudes up to an overall
constant $c_N$.

\subsec{Derivation of  the Veneziano formula}

Example one of the previous section has already
produced one identity on the tachyon scattering amplitude
$\CA(s,t)$. The relation \fdri\ by itself is not sufficiently strong to
determine the function $\CA$.  However, we can combine
it with \fdrvii\  using the  decoupling of
BRST trivial states. Decoupling of the longitudinal
photon implies that $\CA_{0100}(s,t|s,t)=0$. Combining
this with \fdrvii\ we get
\eqn\fdrviii{
s\CA(s-1,t)=t\CA(s,t-1)
}
Now the recursion relations
\fdri\ and \fdrviii\  determine the value of $\CA$ for
$s,t\in\IZ_+$ to be given by the Veneziano formula:
\eqn\venez{
\CA_{0000}=\CA(s,t)=c_4 {\Gamma (s+1)\Gamma(t+1)\over \Gamma(s+t+2)}
}
where $c_4=\CA(0,0)$ is assumed nonzero.

In order to obtain the amplitude for all $s,t$ we must
``analytically continue from the integers,'' an idea familiar
from studies of the $S$-matrix in $D=2$ spacetime
dimensions. (For reviews see
\ref\reviews{P. Ginsparg and G. Moore, ``Lectures
on 2D Gravity and 2D String Theory,'' hep-th/ 9304011, Lectures
given at the 1992 TASI summer school\semi
I.R. Klebanov,
``String theory in two-dimensions,'' hep-th/9108019;
In Trieste 1991, Proceedings , String Theory and quantum gravity\semi
D. Kutasov, ``Some Properties of (Non)Critical
Strings,''  hep-th/9110041; In Trieste 1991, Proceedings ,
String Theory and quantum gravity.}. )
It is at this
point that we must invoke the analyticity properties
{\bf AP1},{\bf AP2}  of section 2.3.

Let us define
\eqn\deficit{
\CA(s,t)=c_4 {\Gamma (s+1)\Gamma(t+1)\over \Gamma(s+t+2)}
\tilde \CA(s,t)
}
The functional equations for $\CA$ imply that
$\tilde \CA(s,t)$ is periodic of period one in
both $s,t$. Moreover, by {\bf AP1} it is an entire function.
By {\bf AP2}, the  entire function
$\tilde \CA$ is  of exponential type
\ref\boas{R. P. Boas,  {\it Entire Functions},
Academic Press, 1954}.
A periodic entire function of exponential type  must be
a trigonometric polynomial \boas, that is:
$$\tilde \CA(s,t)=\sum_{n,m} c_{n,m} e^{2 \pi i(ns + mt)} $$
where the sum is finite.
Now applying {\bf AP2} again
we see that $\tilde \CA$ must in fact be constant,
so $\tilde \CA=1$.

In conclusion, with a mild analyticity assumption we
see that the Veneziano amplitude is fixed by symmetry.

\subsec{n-particle scattering, $n\leq 26$}

The above procedure can be extended to higher
point functions, although the amount of work
involved goes up rapidly with $n$.  We take
the independent kinematic variables to be
$s_{ij}$ for $1\leq i<j\leq n-2$ and $s_{i,n-1}$
for $1\leq i \leq n-3$. The equation
\eqn\snmn{
2-n=\sum_{1\leq i<j\leq n-1} s_{ij}
}
expresses $s_{n-2,n-1}$ in terms of the other
invariants. It is often useful to think of the variables
as an upper triangular matrix:
\eqn\uptrn{
s=\left (\matrix{
* & s_{12}&s_{13}& \cdots & s_{1,n-1} \cr
0 & *& s_{23}& \cdots & s_{2,n-1} \cr
\vdots & & &\ddots & \vdots \cr
0 & \cdots & * & s_{n-3,n-2}&s_{n-3,n-1} \cr
0 & \cdots & & * &s_{n-2,n-1} \cr
0 & \cdots &0 & 0 &* \cr
}\right )
}
where $s_{n-2,n-1}$ is not an independent
variable but is fixed  by \snmn.

The tachyon amplitudes are functions on the
space of upper triangular matrices \uptrn\
defined by \kbnielsen.

The generalization of \fdri\ is
\eqn\trngl{\eqalign{
\CA(\dots, s_{ab} ,\dots ,s_{ac}, \dots, s_{bc}, \dots )=&
\CA(\dots, (s_{ab}-1), \dots, (s_{ac}+1),\dots, s_{bc},\dots )\cr
-\CA(\dots, (s_{ab}-1), \dots, & (s_{ac}),\dots, (s_{bc}+1) , \dots )\cr}
}
which holds
$\forall a,b,c$ such that $  1\leq a<b<c\leq n-1$
where all other variables in the ellipsis are held fixed.  We call
these ``triangle relations'' since they relate a triangle of
variables in \uptrn.

Using the triangle relations one easily reduces an arbitrary
tachyon amplitude for $s_{ij}\in \IZ_+$, $1\leq i<j\leq n-2$ to
the case where:
$$s=\left (\matrix{
* & 0&0& \cdots & 0&s_{1,n-1} \cr
0 & *& 0& \cdots & 0&s_{2,n-1} \cr
\vdots & & &\ddots & 0&\vdots \cr
0 & \cdots & 0 & *&0&s_{n-3,n-1} \cr
0 & \cdots & &0& * &s_{n-2,n-1} \cr
0 & \cdots &0 & 0 &0&* \cr
}\right )
$$
We denote a tachyon amplitude evaluated for
such a set of invariants by the function
$F(s_{1,n-1},\dots s_{n-3,n-1}).$

Next, one may write lightlike relations connecting
$\CA_{0\cdots 0 1 0 \cdots 0}$ to tachyon
scattering. As above,  one can
put $\CA_{0\cdots 0 1 0 \cdots 0}$
to zero by evaluating at special polarization invariants
corresponding to longitudinal photons.
In this way we obtain
a set of $n-1$ relations for the function $F$. These
equations are:
\eqn\ltlike{\eqalign{
0=\sum_{1\leq j\leq n-2: j\not= a}
\Biggl[\sum_{\epsilon\in \IZ_2^{n-2}:\epsilon_j=\epsilon_a=0} &
(-1)^{|\vec \epsilon|} F\biggl(\vec s+\bigl(n-4-|\vec \epsilon|\bigr)\hat e_a+
\vec \epsilon\biggr)\Biggr]\cr
+s_{a,n-1}\sum_{\epsilon\in \IZ_2^{n-2}:\epsilon_a=0} &
(-1)^{|\vec \epsilon|} F\biggl(\vec s+\bigl(n-4-|\vec \epsilon|\bigr)\hat e_a+
\vec \epsilon\biggr)\cr}
}
which holds for $1\leq a\leq n-3$ and where
$\hat e_a$ is a unit vector in the $a$ direction,
$|\vec \epsilon|$ is the sum of the nonzero entries,
and the last entry of an $n-2$ vector is dropped in
the argument of $F$.
We have two additional identities for $a=n-2$:
\eqn\ltlikei{
\eqalign{
(6-2n-\sum_1^{n-3} s_{i,n-1})
\sum_{\epsilon\in \IZ_2^{n-3}}  &
(-1)^{|\vec \epsilon|} F\biggl(\vec s+
\vec \epsilon\biggr)
= \cr
&\sum_{1\leq j\leq n-3}
\Biggl[\sum_{\epsilon\in \IZ_2^{n-3}:\epsilon_j=0}
(-1)^{|\vec \epsilon|} F\biggl(\vec s+
\vec \epsilon\biggr)\Biggr]\cr}
}
and for $a=n-1$:
\eqn\ltlikeii{
\sum_{1\leq j\leq n-3} s_{j,n-1} F(\vec s -\hat e_j)=
(n-3 +\sum_1^{n-3}s_{i,n-1}) F(\vec s)
}

As we will show below,
these equations are sufficient to determine the
functional dependence of $F$ for $s_{ij}\in \IZ$, up
to a finite set of arbitrary constants.
The remaining undetermined constants can
be fixed (up to one overall scale) by  use of the
axiom {\bf AP1}.  We illustrate this by giving the
general solution to the functional equations
\ltlike\ltlikei\ltlikeii\  for
the case of the 5-particle function. We then give
the general argument.

\subsec{5-particle function}

The functional equations become
\eqn\fvpti{\eqalign{
x\Biggl[ F(x+1,y)-F(x,y)-F(x,y+1)&+F(x-1,y+1)  \Biggr]+\cr
+2F(x+1,y) -F(x,y)- & F(x,y+1)=0\cr
y\Biggl[ F(x+1,y)-F(x+1,y-1) & -F(x,y+1)+F(x,y)\Biggr] + \cr
+F(x+1,y) &+F(x,y)- 2F(x,y+1)=0\cr
(x+y+4)\Biggl[ F(x+1,y+1)-F(x+1,y) & -F(x,y+1)+F(x,y)\Biggr] + \cr
F(x+1,y)+ & F(x,y+1)-2F(x,y)=0 \cr
(x+y+2)F(x,y)=x F(x-1,y)&+ yF(x,y-1)\cr}
}

The equations must be evaluated in the
quadrant $x=-2-a,y=-2-b$, $a,b\geq 0$, otherwise
the amplitude might have poles.
The general solution to \fvpti\ in this quadrant is easily found
to be
\eqn\fvptii{
F(x,y)=
{2 \biggl[4 \beta +(3 \alpha-4 \beta)x + (8 \beta-3 \alpha) y)
\biggr]\over  (x+1)(y+1)(x+y+2)}
}
where $\alpha,\beta$ are undetermined. By
{\bf AP1} we only allow poles in $x$ or in
$x+y$ ( since $y$ itself is not related to the squared
momentum in any channel).  This  requirement
fixes $3 \alpha = 4 \beta$ so that
\eqn\fvptiii{
F(x,y)= 6 \alpha
{1 \over  (x+1)(x+y+2)}
}
It is now a simple matter to use the triangle relations
together with the relation
$$\sum_{j=0}^n \left (\matrix{
n\cr
j\cr
}\right ) (-1)^j {1\over n-j+x}= -{\Gamma(-n-x)\Gamma(n+1)\over
\Gamma(1-x)}
$$
to obtain the complete five-particle amplitude
 in terms of a
generalized hypergeometric function:
\eqn\fvptiiii{\eqalign{
\CA(s_{12},s_{13},s_{14},s_{23},s_{24})
=\qquad\qquad \qquad  \qquad\qquad & \qquad\qquad \cr
6 \alpha
{\Gamma(s_{12}+1)\Gamma(s_{23}+1)\Gamma(1+s_{34})
\over \Gamma(2+s_{23}+s_{34})} &
{\Gamma(-1-s_{12}-s_{13}-s_{14})\over
\Gamma(-s_{13}-s_{14})} \cr
{}_3 F_2\biggl(
 -s_{13}, -1-s_{12}- s_{13}-s_{14}, 1+s_{34} ;
  -s_{13}-s_{14}, & 2+s_{23}+s_{34}  \bigl | 1\biggr)\cr}
}
where $s_{34}$ is defined by \snmn.

Equation \fvptiiii\ is derived for integral  values of $s_{ij}$.
In order to ``continue'' to all values of $s_{ij}$
we must combine results on entire functions with
the analyticity property {\bf AP2} of section 2.3, as in
the previous section. This is straightforward when
$s_{13}=0$. In the general case we must use a
Mellin-Barnes representation for ${}_3 F_2$ to
establish appropriate asymptotics.

The formula \fvptiiii\  can actually be derived directly
from the integral representation \kbnielsen\ using
formulae in, e.g.,
\ref\hyper{H. Exton, {\it Handbook of Hypergeometric
Integrals}, John Wiley,  1978}.
Similar  formulae have appeared in a different
context in
\ref\bienk{J. Bienkowska, ``The renormalization group flow in 2D  N=2 SUSY
Landau-Ginsburg models,'' hep-th/9109003.}.

\subsec{$6\leq n\leq 26$}

The above discussion generalizes to $n$- particle
scattering for $6\leq n\leq 26$. The
analyticity property {\bf AP1} combined with the
functional equation \ltlikeii\ is sufficiently
strong to obtain the general formula.

We now prove this. To begin
we use \ltlikeii\ with $s_2=s_3=\cdots =0$:
\eqn\ltkiv{
s_1 F(s_1-1,\vec 0)=(n-3 + s_1) F(s_1,\vec 0)
}
which implies
\eqn\ltkv{
F(s_1,\vec 0) = {c\over (s_1+1)\cdots (s_1+n-3) }
}
where $c$ is a constant. Now given $F(s_1,\vec 0)$ we can
again use \ltlikeii\ to derive
\eqn\ltkvi{
F(s_1,1,\vec 0) = {c_1 s_1 + c_2\over
(s_1+1)\cdots (s_1+n-2) }
}
where $c_1,c_2$ are constants. We can carry on in this way
and easily establish by induction that if $s_2,s_3,\dots $
are nonnegative integers then
\eqn\ltkvii{
F(s_1,s_2,\dots s_{n-3}) = {P^{s_2,\dots s_{n-3}}(s_1)\over
(s_1+1)\cdots (s_1+s_2+\cdots + s_{n-3} + n-3) }
}
where $P^{s_2,\dots s_{n-3}}(s_1)$ is a polynomial
of degree $s_2+\cdots +s_{n-3}$.

Now, as with the 5pt
function we can fix the coefficients of the polynomial
by invoking {\bf AP1}.
By {\bf AP1} the
function $F(\vec s)$ can only have poles when
\eqn\poloc{\eqalign{
s_{1,n-1}&\in \{-1,0,1,2,\dots \} \cr
s_{1,n-1}+s_{2,n-1}&\in \{-2,-1,0,1,2,\dots \} \cr
s_{1,n-1}+s_{2,n-1}+s_{3,n-1}&\in \{-3,-2,-1,0,1,\dots \} \cr
\vdots\qquad\qquad & \qquad\qquad \vdots \cr
s_{1,n-1}+\cdots s_{n-3,n-1} & \in \{-(n-3),-(n-4),\dots\} \cr}
}
Therefore, the polynomial in the numerator of \ltkvii\
must cancel the poles:
\eqn\canpol{\eqalign{
s_{1,n-1}&\in \{-s_2-1, \dots, -2 \} \cr
s_{1,n-1}&\in \{-s_2-3, \dots, -s_2-s_3-2\} \cr
\vdots\quad& \qquad \qquad\qquad \vdots \cr
s_{1,n-1} & \in \{-s_2-\cdots-s_{n-4}-(n-3),\dots,
-s_2-\cdots-s_{n-3}-(n-4)\} \cr}
}
This fixes all the constants in $P$ up to an overall scale
and we obtain the result
\eqn\finres{
F(\vec s)=c_n \prod_{j=1}^{n-3} {-1\over j+\sum_{\ell=1}^j s_{\ell,n-1} }
}
which can also be checked directly from the integral
representation \kbnielsen.

One can proceed from here to evaluate the general
tachyon amplitude using the triangle relations and
{\bf AP2}.  The
result after ``putting back'' $s_{12},s_{23},\dots s_{n-3,n-2}$ is
still a product of gamma functions. The general result is
a  multiple hypergeometric function of $\half (n-3)(n-4)$ arguments.
For example, the six-point function turns out to be:
\eqn\sxpt{\eqalign{
\sum_{j_1,j_2,j_3\geq 0}  {(-s_{14})_{j_1}(-s_{24})_{j_2}(-s_{13})_{j_3}
\over j_1 ! j_2 ! j_3 !}&
{\Gamma(-s_{12}-s_{13}-s_{14}-s_{15}+j_1+j_3)\over
\Gamma(1-s_{13}-s_{14}-s_{15}+j_1+j_3) }\times\cr
 {\Gamma(4-s_{34}-s_{35}-s_{45}+j_1+j_2+j_3)\over
\Gamma(5-s_{23}-s_{34}-s_{35}-s_{45}+j_1+j_2+j_3) }
&{\Gamma(1+s_{12})\Gamma(1+s_{23}) \Gamma(1+s_{34})\over
\Gamma(5-s_{23}-s_{34}-s_{35}-s_{45}+j_1+j_2) }\cr}
}
Multiple hypergeometric functions have been studied to some
extent in the literature, see, e.g., \hyper. In order to
apply {\bf AP2} we must give a Mellin-Barnes representation
to series like \sxpt\  to establish the
appropriate asymptotics in $s_{ij}$. We have not carried
out this procedure in complete detail, but fully expect that it can
be done.

\newsec{$S$ is unique}

We now argue that the solution to the
bracket relations is essentially unique.

\subsec{4-particle $S$-matrix}

We begin by showing that the identities \fdr\
completely fix the
4-particle $S$-matrix for all particles in terms
of the level zero $S$-matrix $\CA(s,t)$.
 Indeed, this can already
be done simply by using the lightlike Ward identities.
The proof of this assertion is a simple application of the
no-ghost theorem and DDF operators
\grschwtt. If $k_0$ is a lightlike
vector, $p_0^2=2$, and $p_0\cdot k_0=1$ then the bracket
$$
\{ i \zeta\c\p X e^{-i\ell k_0 X},\quad \c \quad\}:
\CH[p_0-(n-\ell)k_0,n-\ell]\to
\CH[p_0-n k_0,n]
$$
is equivalent to applying  DDF
operators $\zeta\c A_{-\ell}$ to $\CH[p_0-(n-\ell)k_0,n-\ell]$.
Therefore, the no-ghost theorem implies that
\eqn\onmap{
\{ , \}:\oplus_{1\leq \ell \leq n}  \CH[-\ell k_0,1]\otimes
\CH[p_0-(n-\ell)k_0,n-\ell]\to
\CH[p_0-n k_0,n]
}
is a surjective map.

We would like to proceed as in
the examples of section four using the
Ward identities to reduce the level-numbers of various
states in the amplitude. We use induction on $n_T=\sum n_i$,
the sum of the level numbers of the states in an amplitude.
Consider the
lightlike Ward identities of the form
\eqn\miii{
\pmatrix{n_1&n_2&n_3&n_4\cr
n_1-\ell&n_2-\ell&n_3&n_4\cr
n_1-\ell&n_2&n_3&n_4\cr
n_1-\ell&n_2&n_3&n_4\cr}
}
for $1\leq \ell \leq n_1$.
These identities relate the amplitude
encoded by the first row to amplitudes with smaller
values of $n_T$.
Since \onmap\ is surjective, we can map an {\it arbitrary}
amplitude, that is, an amplitude where the level $n_1$
state has an arbitrary polarization, to amplitudes with
smaller total level number.

\subsec{$N$-particle scattering}

The above argument generalizes easily
to $N$-particle
scattering, using the level matrix
\eqn\nprtcl{
\pmatrix{n_1&n_2&n_3&\dots &n_N\cr
n_1-\ell&n_2-\ell&n_3&\dots& n_N\cr
\vdots& & \ddots &  & \vdots \cr
n_1-\ell&n_2&n_3&\dots &n_N\cr
n_1-\ell&n_2&n_3&\dots &n_N\cr}
}
as long as we can specify the levels and
kinematic invariants arbitrarily.  From the
discussion of section 3.3 we see that this is
possible for $N\leq 26$
and hence $N$-particle scattering
for arbitrary levels can be expressed in
terms of tachyon scattering.

We can complete the argument for uniqueness
by using the results of section 5, where we showed
that the bracket relations determine the $N$-tachyon
amplitude up to a constant  $c_N$ (see \finres).

\subsec{Summary}

We have established:

\noindent
{\bf Theorem 1}:
A multilinear function  $\CA_N:\CH^{\otimes N}\to \IC$, $N\leq 26$,
which satisfies

\noindent
1. Poincar\'e invariance

\noindent
2. The bracket relations \genfdr.

\noindent
3. The analyticity properties {\bf AP1} and {\bf AP2}

\noindent
is uniquely determined up to an overall constant, $c_N$.

Moreover, an easy argument gives:

\noindent
{\bf Theorem 2}:
A set of multilinear functions  $\{ \CA_N \}_{N\leq 26}$
which satisfy 1,2,3 above as well as  {\bf AP3}
are uniquely specified up to one parameter
$\kappa$ by $c_N=\kappa^{N-2}$.

Theorem 2 follows in the standard way by
examining the residue of a tachyon amplitude at
a tachyon pole. By {\bf AP3} this must be a product of
tachyon amplitudes.
Hence $c_{N_1+1} c_{N_2+1}=c_{N_1+N_2}$ so
$c_N=\kappa^{N-2}$, where $\kappa$ is the string
coupling.

\medskip
\noindent
{\bf Remark}:  We offer one speculation on how the above
results can be generalized to $N>26$. In establishing
theorem 1 we only used the bracket relations for currents
$J$ at levels $0,1$. In order to extend the results to
$N>26$ it will be necessary to use higher level currents.
In the notation of section 3.3 , by fixing the $n_i,\tilde n_i$
we are only free to choose momenta to fix independent
invariants $s_{ij}$ ($i\not=j$), on a codimension $N$ subvariety
$\CI(n_i,\tilde n_i)$ of the variety $\CI(n_i)$ of all invariants $s_{ij}$
at fixed $n_i$.
Nonetheless, it is possible that the amplitude can still
be uniquely determined by generalizing the idea of ``analytic
continuation from the integers'' used previously. By
varying the $\tilde n_i$ at fixed $n_i$
one can possibly determine the
amplitude $\CA_{n_1,\dots n_N}$
on ``enough'' subvarieties $\CI(n_i,\tilde n_i)\subset \CI(n_i)$
that, when combined with the axioms {\bf AP1},{\bf AP2},
the amplitude is fixed on the entire variety $\CI(n_i)$.

\newsec{Closed Strings}

The closed string cohomology for
ghost number $(1,1)$ is
\eqn\clscoh{\CH_c=
H_c^{1,1} =\oplus_{n\in \IZ^+} \int_{\IR^{25,1}} dp \quad
\CH[p,n] \otimes \bar{\CH}[p,n]
}
where the superscript refers to left and right
ghost number and the bar always stands for ``right-mover,'' and
not ``complex conjugate.''
S-matrix amplitudes are multilinear functions
$\CA_c:\CH_c^{\otimes n}\to \IC$ constructed as
follows. The operator formalism associates a
measure $\Omega(V_1\otimes \bar{V}_1,\dots,
V_n\otimes \bar{V}_n)$ on the
moduli space of the n-punctured Riemann sphere,
$\CM_{0,n}$.  We integrate $\Omega$ over
$\CM_{0,n}$ for an appropriate domain of
$s_{ij}$ and continue analytically from there.
There are two ways we may try to extend
the above results to closed strings.

\subsec{Factorization of Amplitudes}

A beautiful result of  Kawai, Lewellen, and
Tye
\ref\klt{H. Kawai, D.C. Lewellen, S.-H. H. Tye,
``A relation between tree amplitudes of closed and open
strings,'' Nucl. Phys. {\bf B269}(1986)1}\
states that  the closed and open four-particle
amplitudes are
related by
\foot{We take $\alpha'=2$ for closed string amplitudes.}
\eqn\clsdop{\eqalign{
\CA_c(V_1\otimes \bar{V_1},  V_2\otimes \bar{V_2},&
V_3\otimes \bar{V_3},V_4\otimes \bar{V_4}) \cr
&=-\sin (\pi t)\CA(V_1,V_2,V_3,V_4) \CA(\bar{V_1},
\bar{V_3},\bar{V_2}, \bar{V_4})\cr}
}
At high energy we have
 \eqn\gmthmi{\eqalign{
\CA_c(V_1\otimes \bar{V_1}, V_2\otimes \bar{V_2},&
V_3\otimes \bar{V_3},V_4\otimes \bar{V_4})  \cr
&\sim \CA^{s.p.}(V_1,V_2,V_3,V_4)
\CA^{s.p.}(\bar V_1,\bar V_2,\bar V_3,\bar V_4)\cr}
}
Similar, but more complicated
remarks hold for $n$-particle scattering.
We may combine  \clsdop\ with previous
results to relate scattering
of massive closed string states to
closed string tachyonic scattering.

\subsec{Algebraic structures for closed strings}

It is possible to extend the definition of the
bracket to the closed string, as shown in
\lziii\relres.
Composing this bracket with $b_0^+=b_0+\bar b_0$ we obtain
a map of the physical states to themselves:
$H_c^{1,1}\otimes H_c^{1,1}\to H_c^{1,1}$.
This map
is essentially a tensor product of open string brackets
and does not turn the physical states into a Lie algebra.
Also, one cannot justify the analog of \genfdr. Thus,
the most straightforward generalization of the
previous discussion does not work.

Instead what one can do is consider $\CH_c$ to be
the ``diagonal'' subspace of  the larger space
$$
\tilde \CH_c=\CH_{\rm open}\otimes \bar{\CH}_{\rm open}
$$
defined by equality of left and right momenta: $p=\bar p$.
Amplitudes on the larger space $\tilde \CH_c$ are
only defined for $n$-tuples which are pairwise
mutually local. The physical amplitudes are a subset of
this expanded set.  It is straightforward to write  bracket
relations for this expanded set of amplitudes. Suppose
$J\otimes \bar J$ has momentum $(q,\bar q)$, and
suppose further that
$q+\sum p_i=0$, $q\c p_i \in \IZ$,
$\bar q+\sum \bar p_i=0$, $\bar q\c \bar p_i \in \IZ$.
Then
\eqn\genfdrcl{\eqalign{
\sum_{i,\bar i} (-1)^{q\c p_2+\cdots q\c p_i}
(-1)^{\bar q\c \bar p_2+\cdots \bar q\c \bar p_{\bar i}}&\qquad \qquad\qquad\cr
\CA_c\bigl(V_1\otimes \bar{V}_1,\dots, \{ J, V_i \}\otimes \bar{V}_i,\dots,
& V_{\bar i}\otimes\{ \bar{J} , \bar{V}_{\bar i} \},\dots,
V_n\otimes \bar{V}_n\bigr) =0\cr}
}
Evidently, the previous sections apply to the
left- and right- degrees of freedom separately
and fully determine the expanded set of amplitudes,
hence {\it a fortiori}, the physical ones, at least for
$n\leq 26$.

One implication of these remarks is that the
$\alpha'\to 0$ limit of scattering
amplitudes of gravitons are in principle
completely determined by the bracket
relations. This raises the interesting issue
of the relation of  bracket algebras
and algebras of vector fields.

\newsec{Conclusions}

\subsec{What we did}

We have shown that
the bracket operation on mutually
local BRST classes may be combined with
Lorentz invariance and analyticity to write
an infinite set of functional relations on
string scattering amplitudes.  These
relations are rather restrictive. The central results of this
paper are theorems 1 and 2 of section 6.3
which state that  the bracket relations together
with the analyticity axioms
{\bf AP1,AP2, AP3} of section 2.3
uniquely determine the genus zero
open string $S$-matrix in terms of a single
free parameter,  $\kappa$, the string coupling
constant, at least for $N\leq 26$-particle scattering.

Some readers will be puzzled by our
emphasis on high-energy limits. This
limit might seem irrelevant if the bracket
relations already determine the amplitude for
all energies. The finite-difference
relations are not,
properly speaking, Ward identities
since they relate amplitudes at
different energies. Our chief
concern has been understanding the
symmetries of string theory and
how they are realized in different
backgrounds. Since high energy
limits of the finite-difference relations
look more like Ward-identities they
deserve special attention.

Some readers will object
that this paper contains nothing new.
After all,
the structure constants of the bracket are
just the on-shell three point functions.
By factorization,
a knowledge of all
the three-point functions  in principle
determines the full $S$-matrix.
If one were mainly interested in explicit formulae
for amplitudes the factorization approach
would be impracticable, whereas our
approach could be made quite efficient.
The important point is, however, that the
present discussion clarifies
the fundamental role of the underlying
algebraic structure of the bracket.

\subsec{What we should do}

There is plenty of room for further work.

Clearly the restriction on $N\leq 26$ for
$N$-particle scattering is extremely
unsatisfactory. It is possible that the
result of this paper can be extended
to $N>26$ by combining the
procedure  of ``analytic continuation from
the integers'' with  bracket relations
for states other than lightlike states.
Indeed, in arriving at our result we have
used only a small subset of the entire
set of bracket relations.

There are several possible
generalizations of the present study. These
include:

1. {\it String perturbation theory}.   Unfortunately, it is
far from clear how to extend the
above results to quantum perturbation
theory.  The problem is that a chiral
BRST class  $J$ is not mutually local
with respect to all states. That means
$\{J,\ \c \ \}$ does not commute with an
insertion of the identity operator $\sum |I\rangle \langle I|$
or, more geometrically, contour integrals of $J$ cannot
be pulled around handles.  This problem disappears
for total closed string compactification,
but, curiously, such theories have other difficulties
with loop amplitudes \finite.  For the same reason
extension to mixed open-closed string scattering
is not trivial.

Our attitude towards this problem is that
in this paper we have managed to understand
better the
{\it classical} symmetries of string theory.
It remains to be seen how the symmetries are
realized quantum-mechanically. We are pursuing
some ideas in this direction.

2. {\it Fermionic strings}. It should be
interesting to generalize these
results to superstring amplitudes. We intend to
return to this in a future work.

3. {\it Other backgrounds}.
We have restricted attention to the
background of 26-dimensional
Minkowski space. Bracket
relations should exist for
amplitudes in any background with an uncompactified
Minkowski space component.  If the monodromy of
the internal parts of vertex operators is abelian one
should be able to adjust spacetime momenta to
obtain mutual locality in some situations.  In such
situations the technical
lemma of section 3.3 shows that one could
write bracket relations for $N\leq d$ -particle
scattering where $d$ is the number of uncompactified
dimensions.
The symmetry algebra of mutually local BRST
classes will depend on the background.
It is not clear how effective the
relations will be in other backgrounds.

The bracket relations are very reminiscent of the
``$W_{\infty}$ Ward identities'' which have been
used to obtain amplitudes for string scattering in
$1+1$ dimensional spacetime
\witzwie
\nref\kutmarsei{D. Kutasov, E. Martinec, and
N. Seiberg, ``Ground Rings and
Their Modules in 2D Gravity with $c\leq 1$ Matter'' (hep-th/9111048),
Phys. Lett. B276 (1992) 437.}
\nref\klebpasq{I. Klebanov and A. Pasquinucci,
``Correlation functions from two-dimensional string Ward identities''
(hep-th/9204052) PUPT-1313.}
\nref\berkut{M. Bershadsky and D. Kutasov,
``Scattering of Open and Closed Strings in $1+1$
Dimensions,''  hep-th/9204049, Nucl. Phys.
{\bf B382}(1992)213.}
\refs{\kutmarsei {--} \berkut}.
These techniques made essential use of the
existence of a nontrivial ghost number zero
cohomology, something which is absent in
the critical bosonic string. We hope that
the bracket relations will play a role in
general backgrounds analogous to
the ``$W_{\infty}$ Ward identities''
of $1+1$ dimensional string theory.

Finally, finite difference relations for correlation
functions are known to arise in certain exactly solvable
quantum field theories as well as in studies of
$q$-deformed affine algebras. It would be very interesting
to discover an underlying quantum group symmetry in
the critical bosonic string $S$-matrix.

\subsec{What are we doing?}

Beyond these questions of generalization there
is the much larger question
of exactly what role the bracket should play in
the formulation of string theory.

We believe that the bracket relations are a stringy
expression of spontaneous symmetry breaking.
On physical grounds one expects that symmetries
which connect scattering of states at different
mass levels must be spontaneously broken.
We will say that a current $J(z)$ (or, more
generally, a ghost number 1 BRST class) is
``broken'' if it is not mutually local with respect to
some on-shell state. In the open string the
on-shell condition is simply $Q \psi=0$ so the
only unbroken currents are $\p X^\mu$.
In the closed string statespace - viewed as a
subspace of $H_{\rm open}\otimes \bar{H}_{\rm open}$-
the on-shell condition requires furthermore that
$p_L=p_R$.  According to this terminology
all holomorphic BRST classes are off-shell and
broken with the exception of $c \p X^\mu$ and its conjugate.
Nevertheless, through the bracket relations
these holomorphic classes
constrain the couplings of on-shell particles.

{}From the above point of view, the uncompactified string
contains infinitely many broken symmetry algebras.
These are the sets $\CL=\{ V_i\}$ of mutually
local BRST invariant states in $\CH_{\rm open}$
which are closed under the bracket. We have seen
in section 3.1 that such sets $\CL$ can be given the
structure of a vectorially-graded Lie algebra.
We can further justify the name ``broken symmetry
algebra'' for such sets $\CL$ by noting that
$\CL$ can also be given a Lie algebra structure,
and that this Lie algebra is an algebra of unbroken
symmetries of some
closed string toroidal compactification.
The reason for this is that, if $\{ p_i\}$ is the set of
momenta associated to $\{ V_i\}$ then
$\Gamma=\langle p_i\rangle$ is an even
integral lattice.  ($\Gamma$ may be Euclidean or
Lorentzian.) The lattice is
even since $p_i^2=2-2n_i$ and
integral by mutual locality.
By the Frenkel-Kac construction
\ref\frnkl{I.B. Frenkel and V.G. Kac, ``Basic
Representations of Affine Lie Algebras and Dual
Resonance Models,'' Inv. Math. {\bf 62}(1980)23.}
\ref\gddrd{P. Goddard and D. Olive, ``Algebras, lattices
and strings,'' in {\it Vertex Operators in Mathematics and
Physics, Proceedings of a Conference} eds. J. Lepowsky,
S. Mandelstam, I.M. Singer, Springer;Intl. Jour. Mod.
Phys. {\bf A1}(1986)303.},
we can introduce
cocycle operators to turn the vectorially-graded
bracket algebra of the $\{ V_i \}$ into a true
 Lie algebra structure on
$$\CL_\Gamma=\oplus_{n\geq 0,p\in\Gamma} \CH[p,n]\qquad . $$
To see that this is the unbroken symmetry of a
toroidal compactification recall that
$$
\eqalign{
(\Gamma;0)\oplus (0;\Gamma) \hookrightarrow
 & \{(p_L;p_R)| p_L,p_R\in \Gamma^*, p_L-p_R\in \Gamma\} \cr
&\cong  II^{D,D} \cr}
$$
and the unbroken symmetry associated with this background
is $\CL_\Gamma\oplus \CL_\Gamma$.

It is further natural to consider {\it maximal} sets of
mutually local BRST classes $\{ V_i \}$. The associated
lattices have rank $26$ and are necessarily hyperbolic.
The associated Lie algebras are unbroken symmetries
of totally compactified backgrounds.  Among these
backgrounds there is a distinguished compactification,
namely, compactification on the torus
defined by  $\Gamma_*=II^{25,1}$ in the open case
or on the Narain compactification
for $\Gamma_* = (II^{25,1};0)\oplus (0;II^{25,1})$
in the closed case. Here $II^{25,1}$ is the unique
even self-dual lattice in $\IR^{25,1}$.
 The corresponding algebra $\CL_*=\CL_{\Gamma_*}$
 has been dubbed the
``fake Monster Lie algebra'' by Borcherds
\ref\borcherds{R. Borcherds, ``Generalized Kac-Moody
Algebras,'' J. Algebra {\bf 115}(1988)501; ``Monstrous
moonshine and monstrous lie superalgebras,''
Invent. Math. {\bf 109}(1992)405}
\ref\gebert{R. Gebert, ``Introduction to Vertex Algebras,
Borcherds Algebras, and the Monster Lie Algebra,''
hep-th/9308151; DESY 93-120}.

Based on the analogy with Euclidean
compactifications, which duly reproduces the
Higgs mechanism in the $\alpha'\to 0$ limit
\grschwtt, we may say that the uncompactified
string is a spontaneously broken gauge theory
with the  gauge algebra  $\CL_*$ (in the
open case) and $\CL_*\oplus \CL_*$ (in the closed case)
broken down to
$\IR^{26}$ and $ \IR^{26}\oplus \IR^{26}$
respectively.

At high energies the bracket relations become
Ward identities.
This should be understood as some
kind of stringy high-energy
symmetry restoration which replaces
the analogous notion in spontaneously broken
gauge theory.
(As shown in appendix B, the standard
field-theoretic approach does not
generalize straightforwardly.)
In this sense the above hyperbolic Lie algebras of toroidal
compactification  ``explain'' the linear relations on
high energy amplitudes discussed in \grsslett.

\medskip
\noindent
{\bf Remark}:
The uniqueness of $S$ is a generalization
of Proposition 14 of \finite, with the fake Monster
Lie algebra replaced by the ``vectorially-graded
Lie algebroid of physical states.''  It is interesting
to note that the structure constants of these two
algebraic objects are closely related.
The structure constants of appendix A
are, essentially, analytic continuations of the structure
constants of $\CL_*$. One could say that the bosonic
string $S$-matrix is made out of the structure constants
of $\CL_*$.

\subsec{What we dream of doing}

It is clear from a study of unbroken symmetries in
toroidal compactification that the Lie algebras
$\CL_*$, $\CL_*\oplus \CL_*$, while distinguished,
 are not the full story.
As explained in
\finite\  $\CL_*\oplus \CL_*$ is not a universal symmetry for
closed string
toroidal compactification. We hope that there is
some kind of ``universal'' algebraic structure in
string theory which will replace compact Lie algebras
in the complete formulation of string theory
as a generalization of gauge theory.

In nonabelian gauge theories like the standard model
the on-shell asymptotic states form representations of the
symmetry group. Since the representations do not just involve
the adjoint there is, in general, no algebraic structure
on the space of on-shell states. String theories
are a very interesting class of theories in which
the on-shell states themselves form an algebra.
String theory might some day be regarded as the theory
of symmetry in its purest form: a single symmetry principle
fixes entirely the particle content and the interactions.

\bigskip

\noindent
{\bf Note added}: The earlier version of this paper
erroneously claimed that the lemma of section 3.3
applied to $N$-particle scattering for {\it all} $N$,
not just $N\leq 26$, and
hence  that bracket relations fixed all $N$-particle
amplitudes. I thank E. Witten for pointing out this
error.  I also thank H. Verlinde  for insisting on the point
when I didn't listen.

\bigskip
\centerline{\bf Acknowledgements}

I would like to thank T. Banks for important
encouragement, and B. Lian and G. Zuckerman
for many useful discussions of their work.
I also thank S. Cordes, I. Frenkel, J. Horne,
R. Shankar, and E. Verlinde for discussions.
I thank the Ecole Normale Superieure (Paris) and
CERN for hospitality extended to me while some
of this research was done.
This work is supported by DOE grant DE-AC02-76ER03075,
DOE grant DE-FG02-92ER25121,
and by a Presidential Young Investigator Award.

\appendix{A}{Table of some structure constants for the bracket}

$\bullet$ $\{ p_1^2=2,p_2^2=2\}$:
\eqn\ti{
\{ e^{i p_1\c X},e^{i p_2\c X} \} =
\cases{
0 & for $p_1\c p_2\geq 0$ \cr
e^{i( p_1+p_2)\c X} & for $p_1\c p_2=-1$ \cr
\CS_n(p_1\c X(z))e^{i( p_1+p_2)\c X} & for $p_1\c p_2=-(n+1)$
}
}
where $\CS_n[f(z)]$ is a differential polynomial in
$f(z)$ defined by Taylor expansion:
$$e^{i (f(w)-f(z))}\equiv\sum_{n\in\IZ} (w-z)^n \CS_n[f(z)]$$
The $\CS_n$ are essentially Schur polynomials in the
derivatives of $f$. The first few nonvanishing ones are
\eqn\schur{\eqalign{
\CS_0&=1\cr
\CS_1&=i\p Y\cr
\CS_2&={i\over 2}\p^2 Y-\half (\p Y)^2\cr}
}

$\bullet$ $\{ p_1^2=0,p_2^2=2\}$:
\eqn\comi{\eqalign{
\{ i\zeta_1\p X e^{i p_1 X}, e^{i p_2 X}\}&= \zeta_1\cdot p_2 e^{i(p_1+p_2)X}
\qquad \qquad p_1\cdot p_2=0\cr
&=i(\zeta_1+(p_2\cdot \zeta_1)p_1) \cdot \p X e^{i(p_1+p_2)X}\qquad p_1\cdot
p_2=-1\cr}
}

$\bullet$ $\{ p_1^2=0,p_2^2=0\}$:
\eqn\comiii{\eqalign{
\{ i\zeta_1\cdot \p X e^{i p_1 X}, i \zeta_2\cdot \p X e^{i p_2 X}\}&
= \bigl[\zeta_1 \c \zeta_2-(\zeta_1\c p_2)
(\zeta_2\c p_1)\bigr] e^{i(p_1+p_2)X}\qquad p_1\cdot p_2=+1\cr}
}

\eqn\comii{\eqalign{
\{ i\zeta_1\cdot \p X e^{i p_1 X}, i \zeta_2\cdot \p X e^{i p_2 X}\}&
= i \zeta' \cdot \p X e^{i(p_2+p_1)X}\qquad p_1\cdot p_2=0\cr
\zeta'=
(\zeta_1\cdot \zeta_2 & -(\zeta_2\cdot p_1)(\zeta_1\cdot p_2))p_1+
(\zeta_1\cdot p_2) \zeta_2 - (\zeta_2\cdot p_1) \zeta_1\cr}
}

$\bullet$ $\{ p_1^2=-2,p_2^2=2\}$:
\eqn\comiv{\eqalign{
\{ V_{\zeta,p_1},  e^{i p_2 X}\}
&=-(p_2+p_1)\c \zeta\c p_2 e^{i(p_2+p_1)\c X}\qquad p_1\c p_2=+1\cr
&= V_{\zeta',p_2+p_1}\qquad\qquad  p_1\cdot p_2=-1\cr
\zeta'=
\zeta +&\bigl(p_1\otimes p_2\c \zeta+\zeta\c p_2\otimes p_1\bigr)
+\half \bigl[(p_2+p_1)\c \zeta\c p_2\bigr]p_1\otimes p_1\cr}
}

\appendix{B}{High energy symmetries and spontaneous
symmetry breaking}

In \finite\ we proposed that the infinite dimensional
symmetries arising upon time-compactification - of
which $\CL_*\oplus \CL_*$  is a spectacular
example - are the high-energy symmetries of
 \grsslett. The idea was that Ward identities
derived in a symmetric compactification could be
``parallel transported'' to the decompactified
background. The decompactification forces momenta
to infinity, hence the connection with high energy
scattering.
This proposal was based on a field
theoretic intuition: in field theory the effects of
spontaneous symmetry breaking disappear at high
energies. It turns out that this intuition is not
very good in string theory. Closed string scattering at
high energy is sensitive to the breaking of toroidal
symmetries. In this appendix we demonstrate that
surprising fact with a few simple examples.

\subsec{Recovery of Ward Identities in Field Theory}

We consider enhanced symmetries and symmetry
breaking in the framework of toroidal compactification.
(We only consider compactification of spacelike
dimensions here.)
Suppose we have an unbroken symmetry current $J_q=e^{i q\cdot Y}$
at a Narain lattice $\Gamma$ which has a vector $(q;0)$ with
$q^2=2$. Consider a theory at $\Gamma' = g(\lambda)\cdot \Gamma$
where $g(\lambda)$ is a one-parameter
family of $O(d,d)$ rotations taking
$(q;0)$ to $(q_L;q_R)$ with $q_R\not=0$.
The symmetry associated with $J$ is spontaneously broken in
spacetime leading to massive gauge bosons with
 mass $m^2_V/m_{pl}^2= q_R^2(\lambda)$ and
$\alpha'\sim 1/m_{pl}^2$ is related to the Planck mass.

Consider  the scattering amplitudes $\CA(V_1,\dots V_n)$
for  (massless and massive) gauge bosons.
The field-theory limit is obtained by letting
$q_R^2,p_i\cdot p_j\to 0$ and expanding in these variables.
The amplitudes have an expansion
$\CA =\sum \epsilon^n \CA^{(n)} $
in powers of $\epsilon=1/m_{pl}$, where each
$\CA^{(n)}$ is a rational function of its arguments.

If we have a Ward identity in the unbroken phase
$\sum \CA(V_1,\dots \delta V_i,\dots V_m)=0$
then we may ask what happens to the
corresponding amplitudes
in the broken phase. The precise definition of
``corresponding amplitudes'' requires a transport
operator $T^\lambda$ on the BRST cohomology. We will simply
rotate momenta by $O(d,d)$ transformations: this
will suffice for our examples.
\foot{See \finite\ for a detailed discussion of a
parallel transport on the entire CFT statespace.}
For each $n$ we have
\eqn\rcwi{
\lim_{\lambda\to 0}\sum_i \CA^{(n)}
(T^\lambda V_1,\dots T^\lambda \delta V_i,\dots T^\lambda V_m)
= 0}
and since $\CA^{(n)}$ are {\it rational} functions of
$p_i\cdot p_j$ and of the symmetry-breaking masses
$q_R^2(\lambda)$ we conclude
that the sum of the terms is proportional to the
symmetry breaking masses. Therefore, for dimensional
reasons, at each order in the expansion in $1/m_{pl}$
the sum of the terms grows more slowly in energy than
the individual terms. In particular, for the leading
terms $\sim m_{pl}^n$ where $n$ is a nonnegative power,
the sum of the terms vanishes. This is the recovery of
the Ward identity at high energy.

{\bf Example}. To be explicit, consider
a circular compactification of a single dimension $Y$.
Let $R=e^\lambda$. At $\lambda=0$ we have an
unbroken $SU(2)\times SU(2)$ gauge theory which is
broken to $U(1)\times U(1)$ for $\lambda>0$.
Introduce the notation:
\eqn\expos{
e(j,\bar j,R)\equiv e^{{i\over \sqrt{2}}({j+\bar j\over R}+(j-\bar j)R)Y}(z)
e^{{i\over \sqrt{2}}({j+\bar j\over R}-(j-\bar j)R)\bar Y}(\bar z)
}
Focus on a single factor $SU(2)\to U(1)$.
In the broken phase we have massive
gauge bosons, call them $W^\pm(\epsilon)$, with corresponding
vertex operators:
\eqn\gbi{W^\pm(\epsilon,p)=
 e(\pm1,0,R)\epsilon\cdot\pb X e^{i p\cdot X}
}
where $X,p$ refer to uncompactified dimensions.
\gbi\ are on shell for
$ p\cdot \epsilon=0, p^2 + 2 \sinh^2 \lambda=0$. There are
also massless gauge bosons for the unbroken $U(1)$ generators,
call them $Z(\epsilon)$ described by vertex operators:
\eqn\gbii{
Z(\epsilon,p)=\cosh \lambda \p Y \epsilon\cdot \pb X e^{i p\cdot X}
}
which are on-shell for $p\cdot \epsilon=0, p^2 =0$.
(The factor of $\cosh \lambda$ in \gbii\ is motivated
by the transport in \finite. It can be dropped without
changing our conclusions.)

At the $su(2)$ point $\lambda=0$ we have the
$su(2)$ Ward identity:
\eqn\sutwi{
\eqalign{
0=\CA\bigl[W^+(\epsilon_1),&Z(\epsilon_2),
W^-(\epsilon_3),Z(\epsilon_4)\bigr](s,u)\cr
+&
\CA\bigl[W^+(\epsilon_1),W^-(\epsilon_2),
Z(\epsilon_3),Z(\epsilon_4)\bigr](s,u)\cr
&\qquad
+\CA\bigl[W^+(\epsilon_1),W^-(\epsilon_2),
W^-(\epsilon_3),W^+(\epsilon_4)\bigr](s,u)\cr}
}

What happens at $\lambda>0$?
For simplicity assume that all polarizations
$\epsilon_i$
are transverse to the plane of scattering and that
$\epsilon_1\c \epsilon_3 \epsilon_2\c \epsilon_4=
\epsilon_1\c \epsilon_4 \epsilon_2\c \epsilon_3=0$,
$\epsilon_1\c \epsilon_2 \epsilon_3\c \epsilon_4=1$.
In the field theory limit we have
\eqn\ftlimt{
\eqalign{
\CA\bigl[W^+,Z,
W^-,Z\bigr](s,u)&= {2(u-m^2)\over s}\cr
&+{1\over m_{pl}^2} (2 m^2 u/s + u -s -2 m^4/s - m^2)+\cdots\cr
\CA\bigl[W^+,W^-,
Z,Z\bigr](s,u)&=
2
+{1\over m_{pl}^2} {(2s-u)(s-m^2)-u^2\over s-m^2}+\cdots\cr
\CA\bigl[W^+,W^-,
W^-,W^+\bigr](s,u)&=2{(2 m^2-s-u)\over s-m^2}\cr
&+{1\over m_{pl}^2}{(u-s)(s+u-2m^2)\over s-m^2}+\cdots\cr}
}
where $s\equiv p_1\cdot p_2, u\equiv p_1\cdot p_3,m^2\equiv 2 \sinh^2 \lambda$.
The sum of the three amplitudes is
$$2 {m^2\over s} {m^2-u\over s-m^2} + {1\over m_{pl}^2}
{m^4\over s}{2m^2-s-2u\over s-m^2}+\cdots$$

The high-energy {\it field-theoretic} limit is now
obtained by letting $s\to -\infty, u\to \infty$ holding
$m^2$ fixed. As promised, at each order in $1/m_{pl}$
the sum vanishes more rapidly than each of the separate
amplitudes, and, in particular, the leading term of
order $\CO(m_{pl}^0)$ vanishes.

\subsec{String theory Ward identities are not recovered at high energy}

We illustrate this with two simple examples.

{\bf Example 1}:

We now take the high energy limit of the three amplitudes
in \sutwi.  The high energy asymptotics of the amplitudes
are:

\eqn\strlimt{
\eqalign{
\CA\bigl[W^+(\epsilon_1),Z(\epsilon_2),
W^-(\epsilon_3),Z(\epsilon_4)\bigr]&\sim
-U^2 (1+\sinh^2 \lambda) e^{4\sinh^2 \lambda\log(-s/u-1)}\cr
{(s+u)^3\over s^3 u^2}[s^2 +s u + 2  u^2 &+ 2\sinh^2 \lambda
u^2](1+\CO(1/s,1/u,1/(s+u))\cr
\CA\bigl[W^+(\epsilon_1),W^-(\epsilon_2),
Z(\epsilon_3),Z(\epsilon_4)\bigr]&\sim
-U^2 (1+\sinh^2 \lambda) e^{4\sinh^2 \lambda\log(u/s+1)}\cr
{(s+u)^3\over s^4 u}[u^2 +u s + 2  s^2 &+ 2\sinh^2 \lambda
s^2](1+\CO(1/s,1/u,1/(s+u))\cr
\CA\bigl[W^+(\epsilon_1),W^-(\epsilon_2),
W^-(\epsilon_3),W^+(\epsilon_4)\bigr]&\sim
U^2 e^{4\sinh^2 \lambda\log[(u/s+1)(-s/u-1)]}\cr
{(s+u)^6\over s^4 u^2} &(1+\CO(1/s,1/u,1/(s+u))\cr}
}
where $U$ is defined in \gmfactor.

While the sum of the three terms in \strlimt\
vanishes at $\lambda=0$ this
is not the case for any $\lambda\not=0$: the sum is not
any smaller in magnitude than any of the three terms in
the sum.

{\bf Example 2}:

At the $SU(2)$ point the vertex operators
$\phi^{\pm\pm}=e(\pm\half,\pm\half,R=1)e^{ip\c X}$
describe a scalar field multiplet in the $(2,2)$ of $SU(2)\times SU(2)$.
At the $SU(2)$ point we have
the Ward identity (again for $J^+$):
\eqn\scwi{
\CA(\phi^{+-},\phi^{-+},\phi^{+-},\phi^{-+})
+\CA(\phi^{--},\phi^{++},\phi^{+-},\phi^{-+})
+\CA(\phi^{--},\phi^{-+},\phi^{+-},\phi^{++})=0
}
In the broken phase $\lambda>0$ and in the high energy limit
the sum of the three terms becomes:
\eqn\scwii{
\biggl[{1\over |z_0|^{R^2}}|z_0-1|^{R^2}
+{1\over |z_0|^{1/R^2}}-1 \biggr] U^2=
\biggl[(\sin^2 \half \theta)^{R^2}
+(\cos^2\half\theta)^{1/R^2}-1 \biggr] U^2
}
Again we see that the sum of terms is no smaller than the
individual terms.

\subsec{Physical interpretation}

We believe that the failure to recover the Ward identities
in the high energy limit
is related to the notorious delocalization of strings at
ultrahigh energies
\ref\delocalize{See, e.g., D. Gross, ``Superstrings and Unification,''
Talk at the XXIV Int. Conf. on High Energy Physics,
Munich, 1988\semi
L. Susskind, ``String theory and the principle of
black hole complementarity,'' hep-th/9307168}.
This delocalization makes the
string amplitudes once again sensitive to relatively
long-distance physics, like the specific nature of the
compactification scheme.

\listrefs

\bye